\documentstyle[12pt,bezier]{article}
\title{\mbox{}\\ \mbox{}\\ Spectroscopy and Renormalisation Group Flow
    of a Lattice Nambu--Jona-Lasinio Model}
\author{A. Ali Khan$^a$, M. G\"ockeler$^{a,b}$, R. Horsley$^{a,c}$, \and
    P.E.L.  Rakow$^d$, G. Schierholz$^{a,c}$ and H. St\"uben$^{d,e}$}
\date{$^a$ Gruppe Theorie der Elementarteilchen,\\
     H\"ochstleistungsrechenzentrum HLRZ, \\
     c/o Forschungszentrum J\"ulich, \\
     Postfach 1913, D-52425 J\"ulich, Germany\\[0.5em]
$^b$ Institut f\"ur Theoretische Physik, RWTH Aachen, \\
     Sommerfeldstra{\ss}e, D-52056 Aachen, Germany\\[0.5em]
$^c$ Deutsches Elektronen-Synchrotron DESY, \\
     Notkestra{\ss}e 85, D-22603 Hamburg, Germany \\[0.5em]
$^d$ Institut f\"ur Theoretische Physik, Freie Universit\"at Berlin, \\
     Arnimallee 14, D-14195 Berlin, Germany\\[0.5em]
$^e$  Address after 1 January 1994: \\
 Konrad-Zuse-Zentrum f\"ur Informationstechnik Berlin, \\
  Heilbronner Stra\ss{}e 10, D-10711 Berlin, Germany\\[2em]
     }
\newcommand{\CHI}{\langle \bar{\chi} \chi \rangle}  % <chibar-chi>
\newcommand{\mR}{\mu_{\rm R}}               % renormalised fermion mass
\newcommand{\fpi}{f_{\pi}}                  % pion decay constant
\newcommand{\Epi}{E_\pi^{(0)}}              % pion ground state energy
\newcommand{\mod}{\mathop{\mbox{mod}}}         % modulo operator
\renewcommand{\Im}{\mathop{\mbox{Im}}}         % Imaginary Part
\renewcommand{\ln}{\mathop{\mbox{ln}}}         % natural logarithm
\newcommand{\MC}{Monte Carlo}
\newcommand{\SD}{Schwin\-ger-Dyson}
\newcommand{\SDeq}{Schwin\-ger-Dyson equation}
\newcommand{\SDeqs}{Schwin\-ger-Dyson equations}
\newcommand{\FS}{finite size}
\newcommand{\PI}{$\pi$}        % pi particle
\newcommand{\SIGMA}{$\sigma$}  % sigma particle
\newcommand{\RHO}{$\rho$}      % rho particle
\newcommand{\aaa}{{\rm(a)}}
\newcommand{\bbb}{{\rm(b)}}
\newcommand{\Fig}[1]{fig.~\ref{#1}}               % Usage: \Fig{label}
\newcommand{\Tab}[1]{table~\ref{#1}}              % Usage: \Tab{label}
\newcommand{\Eq}[1]{eq.~(\ref{#1})}               % Usage: \Eq{label}
\newcommand{\CAPTION}[2]{\caption[#1]{\sl #2} \label{#1}}% \CAPTION{label}{text}
\newcommand{\FRAME}[1]{}
\newcommand{\PSdir}{}
\newcommand{\PSinput}[1]{\includegraphics{#1}}
\newcommand{\PSfigure}[6]{\begin{figure}[#1]%
\begin{center}%
\setlength{\unitlength}{1cm}%
\begin{picture}#2#3\FRAME{#2}%
\PSinput{\PSdir#5.ps}%
\end{picture}%
\end{center}%
\CAPTION{#4}{#6}%
\end{figure}}%
\newcommand{\Hfig}[3]{\PSfigure{#1}{(12.0,13.5)}{(0,1.6)}{#2}{#2}{#3}}
\newcommand{\ifig}[3]{\PSfigure{#1}{(10.0,12.0)}{(0,1)}{#2}{#2}{#3}}
\newcommand{\hfig}[3]{\PSfigure{#1}{(14.1,8.0)}{(0,1)}{#2}{#2}{#3}}

\advance\textheight by \topskip
\makeatletter
\@addtoreset{equation}{section}
\makeatother

\begin{document}
\maketitle
\begin{abstract}
We investigate a lattice Nambu--Jona-Lasinio model both by the Monte
Carlo method and Schwinger-Dyson equations. A comparison allows the
discussion of finite size effects and the extrapolation to infinite
volume. We pay special attention to the identification of particles
and resonances. This enables us to discuss renormalisation group flows
in the neighbourhood of the critical coupling where the chiral
symmetry breaking phase transition takes place. In no region of the
bare parameter space do we find renormalisability for the model.
\end{abstract}
\noindent
\vspace*{-48\baselineskip}

\noindent
{\tt
DESY 93-195 \hfill ISSN 0418-9833\\
FUB-HEP/93-14 \\
HLRZ-93-51 \\
December 1993
}
\thispagestyle{empty}
\newpage

\section{Introduction}

The study of four fermion interactions has a long history. Fermi originally
introduced the idea as a theory for weak interactions~\cite{Fermi}. Later Nambu
and Jona-Lasinio proposed their model~\cite{Nam61} (in the following called the
NJL model) to describe spontaneous chiral symmetry breaking and dynamical mass
generation. This was one of the first works to note the association of a
massless Nambu-Goldstone boson with the spontaneous breaking of a continuous
symmetry~\cite{Gold61}.

 Over the years there have been efforts to investigate
the vacuum structure of strongly interacting four fermion
theories~\cite{others}. Recently there has been renewed interest in this
subject in connection with technicolour and the top-mode standard
model~\cite{technicolour,top-mode}. For these applications one is mainly
interested in the physics of the broken phase where the fermion mass is
 dynamically generated.
 Additional interest was generated by the
discovery that a generalized NJL model is equivalent to the Higgs-Yukawa
model~\cite{Wil73,Higgs} and possibly to $SU(N)$ gauge theory~\cite{SU(N)}.
Furthermore there have
been various investigations of the gauged NJL model~\cite{gauged-NJL} which
show a  line of second order phase  transitions linking
the pure NJL model and QED. This
raises the question of whether the two models fall into the same universality
class. Our investigation of QED~\cite{Goc92} has shown that in the
strong coupling regime (including
 the critical region) it is impossible to vary the cutoff without changing
the physics, while in the weakly coupled region this is possible.
 We have called such
 behaviour weak renormalisability.

 In this paper we are mainly
 interested in
the model's phase transition and critical behaviour. Since this falls into
 the large
coupling region, nonperturbative methods are required.
 We use the lattice \MC\
simulation and the semi-analytic \SD\ method~\cite{S-D,Roman}.
The \MC\ method has the advantage of being exact but can only be applied on
finite lattice volumes. The \SD\ approach on the other hand can be extended
to infinite volumes but suffers from having an uncontrolled approximation. The
synergy of both methods allows us to go much further than with either method
alone.

We have looked at the chiral condensate in order to map out the phase
diagram. We find a second order chiral phase transition implying
infinite correlation length which in
principle allows us to obtain a continuum theory. In the vicinity of the
critical point we calculate the renormalised fermion mass ($\mR$) and
the energy levels of the pseudoscalar (\PI, Goldstone boson),
scalar (\SIGMA) and
vector (\RHO) fermion-antifermion composite states as well as the
pion decay constant ($\fpi$).

A major problem is the identification of resonances and bound states on
a finite lattice. For the \PI\ and \RHO\ we were able to identify the
first excited energy levels by our \MC\ calculation. These compare favourably
with the results from the \SDeqs. Thus we feel encouraged to take the
thermodynamic limit of the \SDeqs. This allows the calculation of the
spectral functions and thus the identification of resonances. With this
identification at hand we construct flow diagrams of dimensionless ratios
of physical quantities. We use them to investigate the renormalisation
group flow in the vicinity of the critical point.

The paper is organized as follows. Our lattice model is defined in section 2.1.
Details of the \MC\ method are given in sections 2.2, 3.1, 3.2 and 4.1 (see
 also~\cite{thesis}). The \SDeqs\ are discussed in
 sections 2.3, 3.2, 3.3, 4.2, 4.3
and the appendices. Our main results of both methods are given in section 5 on
meson spectroscopy and in section 6 for the renormalisation group flows.
Finally in section 7 we present our conclusions.
\newpage

\section{The Model}

Because we are interested in chiral symmetry properties we use staggered
fermions throughout.

\subsection{The Action}

The NJL model is defined by the continuum action
\begin{equation}
 S = \int d^4x\, \left\{ \bar{\psi}(x) \gamma_{\mu} \partial_{\mu} \psi(x)
  -g_0 \left[
  \left(\bar{\psi}(x)\psi(x)\right)^2
  - \left(\bar{\psi}(x) \gamma_5 \psi(x)\right)^2 \right] \right\}
\end{equation}
which is chirally invariant.
For staggered fermions we take the action
\begin{eqnarray}
 S & = & \frac{1}{2} \sum_{x, \mu} \eta_{\mu}(x)
     \left[ \bar{\chi}(x) \chi(x + \hat{\mu})
         - \bar{\chi}(x + \hat{\mu}) \chi(x) \right]
   + m_0 \sum_x \bar{\chi}(x) \chi(x) \nonumber \\
   & - & g_0 \sum_{x, \mu} \bar{\chi}(x) \chi(x)
                     \bar{\chi}(x + \hat{\mu}) \chi(x + \hat{\mu})
                     \label{Eq-S},
\end{eqnarray}
where
\begin{equation}
 \eta_{\mu}(x) = (-1)^{x_1 + \cdots + x_{\mu - 1}},\; \eta_1(x) = 1.
\label{eta-defn}
\end{equation}
The interaction is chosen as the simplest one that is invariant under
the continuous chiral transformation
\begin{equation}
 \chi(x) \rightarrow e^{i \alpha \epsilon(x)} \chi(x),\;
 \bar{\chi}(x) \rightarrow e^{i \alpha \epsilon(x)} \bar{\chi}(x),
\end{equation}
with
\begin{equation}
 \epsilon(x) = (-1)^{x_1 + x_2 + x_3 + x_4}.
\end{equation}
The mass term acts as a chiral symmetry breaking external source and as an
infrared regulator.

\subsection{\MC\ Simulation}

In order to use the \MC\ method we first have to integrate out the
fermion fields. Therefore we have to rewrite
the interaction in a bilinear form. This is done by
introducing an auxiliary field
 $\theta_{\mu}(x) \in [0, 2 \pi)$~\cite{Lee87,Booth}:
\begin{equation}
 S_{\rm int} = \sqrt{g_0} \sum_{x, \mu} \eta_{\mu}(x)
   \left[ \bar{\chi}(x) e^{i \theta_{\mu}(x)} \chi(x + \hat{\mu}) -
   \bar{\chi}(x + \hat{\mu}) e^{-i \theta_{\mu}(x)} \chi(x) \right].
\end{equation}
The Hybrid Monte Carlo method was employed to generate the configurations.
Lattices of size $V = L_s^3 \times L_t = 6^3 \times 12$, $8^4$,
$8^3 \times 16$,
$8^3 \times 32$ and $12^4$ were used.
Boundary conditions were periodic in the spatial directions and
anti-periodic in the time direction.  We investigated the model for $m_0
= 0.01$ to 0.09 and $g_0 = 0.21$ to 0.32. For each parameter set 1000
trajectories were generated for measurements. The trajectory length was
always $\approx$~0.8. The number of time steps was adjusted such that acceptance
rates between 70 and 80\% were achieved. Propagators were calculated
with the conjugate gradient method. As convergence criterion
we required that the norm of the rest vector was smaller than
$10^{-5}$. The chiral condensate was
calculated with a stochastic estimator (see, e.~g.,~\cite{Bit89}).

\subsection{\SD\ Equations}

The model (\ref{Eq-S}) was also studied by solving truncated \SDeqs\
up to order $g_0$ and $g_0^2$ on the lattice. To make a direct
comparison with \MC\ results possible we use the same action
 with the same cut-off (for another application of \SDeqs\ to
 a lattice theory see~\cite{Sloan}).

 In order
to write the equations in this section in a compact form
it is useful to give the lattice action (\ref{Eq-S}) with a
 slightly different notation:
\begin{equation}
  S  =
    \sum_{\mu}  \bar{\chi}_x
        d^{\mu}{}^x_y
      \chi^y
        + m_0 \bar{\chi}_x \chi^x
    -  g_0  \bar{\chi}_x \chi^{x'}
            c^{x\ y}_{x'y'}
                     \bar{\chi}_y \chi^{y'} \,.
                     \label{SumAction}
\end{equation}
 The fields at site $x$ are $\bar{\chi}_x$ and $\chi^x$, and the
 Einstein summation convention is to be used whenever a site index
 occurs twice. The meaning of the kinetic matrices $d^\mu$ and of the
 coupling tensor $c^{x\ y}_{x'y'}$ should be clear by comparison with
 (\ref{Eq-S}):
  $d^{\mu\ }{}^x_y = +\frac{1}{2}\eta_\mu(x)$ if $y=x+\hat{\mu}$,
  $d^{\mu\ }{}^x_y = -\frac{1}{2}\eta_\mu(x)$ if $y=x-\hat{\mu}$
  and zero otherwise;
 $c^{x\ y}_{x'y'}=1$ if $x=x'$, $y=y'$ and $x$ and $y$ are neighbours,
 and is zero otherwise. It is convenient to give the coupling tensor
 four indices so that we can use the Einstein summation convention
 and to remind us that more general four-fermi coupling terms are
 possible.

 The lattice Feynman rules
 are that the bare staggered fermion propagator is
 \begin{equation}
   G^{\rm bare} =  \left( m_0 \delta^x_y
     + \sum_{\mu} d^{\mu\ }{}^x_y \right)^{-1} \,,
  \label{freeprop}
\end{equation}
and the vertex is
\begin{equation}
  - g_0 c^{x\ y}_{x'y'} \,.
\end{equation}
When the interaction is turned on the bare fermion propagator
 is replaced by the full fermion propagator
\begin{equation}
  G^{\rm full} =
  \left( M^x_y
     + \sum_{\mu} D^{\mu\ }{}^x_y \right)^{-1} \,.
\end{equation}
  The new matrices $M^x_y$ and $D^{\mu\ }{}^x_y$  are generalisations
of the matrices $m_0 \delta^x_y$ and $d^{\mu\ }{}^x_y$ appearing in
(\ref{freeprop}) and have the same symmetries, e.~g., $M^x_y$
is symmetric under reflections and links sites with even
separations in violation of chiral symmetry.
On the other hand $D^{\mu\ }{}^x_y$
is odd in the $\mu$ direction and even in the other directions
and only links sites with odd separations and so respects
chiral symmetry. $M$ is invariant under translations,
$D^\mu$ acquires the same factors of $\pm 1$ as $d^\mu$.

\PSfigure{t}{(14.1,9)}{(0,1)}{fig-sd-g}{fig-sd-g}{Pictorial
representation of the \SDeq\ for the fermion propagator.}

In \Fig{fig-sd-g} we show graphically the \SDeqs\ for the fermion
propagator.
It is not possible to solve the full \SDeqs.
They have to be truncated at some stage.
The crudest approximation is to keep only term (a) which gives the
gap equation, namely
\begin{eqnarray}
 D^{\mu\ }{}^x_y &=& d^{\mu\ }{}^x_y \,, \nonumber \\
 M^x_y &=& N \delta^x_y \,,
\end{eqnarray}
where $N$ satisfies
\begin{equation}
   N = m_0 + 8 g_0  \frac{1}{V} \sum_{k}
          \frac{N}{N^2 +\sum_\nu \sin^2k_\nu} \,.
\end{equation}
The $k$-sum runs over all momenta consistent with the boundary conditions.
The fermion mass $\mR$ is given by the position of the pole and is $\mR =
\sinh^{-1} N$.

The next approximation is to keep terms (a) and (b).
Term (a) leads simply to a renormalisation of the mass and term (b) leads to a
renormalisation of the kinetic energy.
We shall call these
the order $g_0$ \SDeqs\ because the error is of
order $g_0^2$. Explicitly they read
\begin{eqnarray}
 D^{\mu\ }{}^x_y &=& F_\mu d^{\mu\ }{}^x_y \,, \nonumber \\
 M^x_y &=& N \delta^x_y \,,
\end{eqnarray}
where $N$ and $F$ satisfy
\begin{eqnarray}
   N &=& m_0 + 8 g_0  \frac{1}{V} \sum_{k}
  \frac{N}{ N^2 +\sum_\nu F_\nu^2 \sin^2 k_\nu }  \,, \nonumber \\
   F_\mu &=& 1 + 2 g_0  \frac{1}{V} \sum_{k}
      \frac{ F_\mu \sin^2 k_\mu}
     {N^2 +\sum_\nu F_\nu^2 \sin^2 k_\nu} \,.
 \label{OGfermi}
\end{eqnarray}
The pole position giving the fermion mass is
\begin{equation}
 \mR = \sinh^{-1} \frac{N}{F_t} \,. \label{eq-mR}
\end{equation}
  In infinite volumes (when the sum is replaced by an integral) $F$
needs no direction index, as all the $F_\mu$'s are identical. However
in a finite system with different lattice lengths and boundary
 conditions in the space and time directions an index is needed.

The order $g_0^2$ equations also include term (c), but with the
full vertex replaced by the bare vertex, an approximation which
introduces an error of $O(g_0^3)$.
%As can be seen from \Fig{fig-sd-g} at this order the full vertex
% can be replaced by the bare vertex.
The resulting equations (given in configuration space) are
\begin{eqnarray}
 M^x_y +\sum_\mu D^{\mu\ }{}^x_y &=&
 m_0 \delta^x_y +\sum_\mu d^{\mu\ }{}^x_y \nonumber \\
  &+& g_0 c^{xu}_{yv} G^{\rm full\ }{}^v_u
    - g_0 c^{xu}_{vy} G^{\rm full\ }{}^v_u \nonumber \\
  &-& g_0^2 c^{xu}_{vr}
          G^{\rm full\ }{}^r_s
          G^{\rm full\ }{}^z_u
          G^{\rm full\ }{}^v_w
          c^{sw}_{zy}            \nonumber \\
  &+& g_0^2 c^{xu}_{vr}
          G^{\rm full\ }{}^r_w
          G^{\rm full\ }{}^z_u
          G^{\rm full\ }{}^v_s
          c^{sw}_{zy} \,.
\end{eqnarray}
 $M$ and $D$ can be projected out by using their different
 behaviours under reflection.
 For the first time the unknown matrices $M$ and $D$ are not simply
 renormalised versions of the bare matrices, but extend further
 in space. Mass and wave-function normalisation become momentum
 dependent.
 The matrices $M^x_y$ and $D^x_y$ are stored in position space.
 These individual components can be thought of as the Fourier
 coefficients for the momentum space  fermion propagator.

The equations were solved numerically by simple iterations.
Typically about 7 iterations were needed in the order $g_0$ case to obtain
convergence and 20--40 iterations in the order $g_0^2$ case.

\clearpage

\section{Chiral Condensate and Fermion Mass}

\subsection{Phase Diagram}

\noindent To find the phase diagram we have computed the chiral
 condensate $\CHI$ also referred to as $\sigma$.
\MC\ results are collected in tables \ref{tab-CHI-11} and \ref{tab-CHI-12}.
To show that we are working in the
vicinity of the critical point we look at a Fisher
plot~\cite{Kou64} of our data, i.~e., we have plotted $\CHI^2$
against $m_0 / \CHI$ in \Fig{fig-fisher}.
The lines connect points of constant $g_0$. If the theory
was mean-field like, these would be straight parallel lines~\cite{Kou64}. We
see that this is approximately the case for our data.
For large $g_0$ the extrapolated lines end on the
$y$-axis giving a finite value for $\CHI$ at $m_0 = 0$,
while for small $g_0$ the
extrapolated lines end on the $x$-axis giving zero for $\CHI$ at $m_0 = 0$.
The line that
corresponds to the critical coupling would end at the origin as is indicated by
the dashed lines in \Fig{fig-fisher}. On the $8^3 \times 16$ lattice this gives
an estimate of about $0.25 < g_c < 0.26$ while on the $12^4$ lattice $g_c$ is
somewhat larger than 0.27. We see from this that there are \FS\ effects in the
data as can also be inferred from a cursory look at
tables \ref{tab-CHI-11} and \ref{tab-CHI-12}. We shall in the following
call the region $g_0 < g_c$ the symmetric phase and the region $g_0 > g_c$
the broken phase.

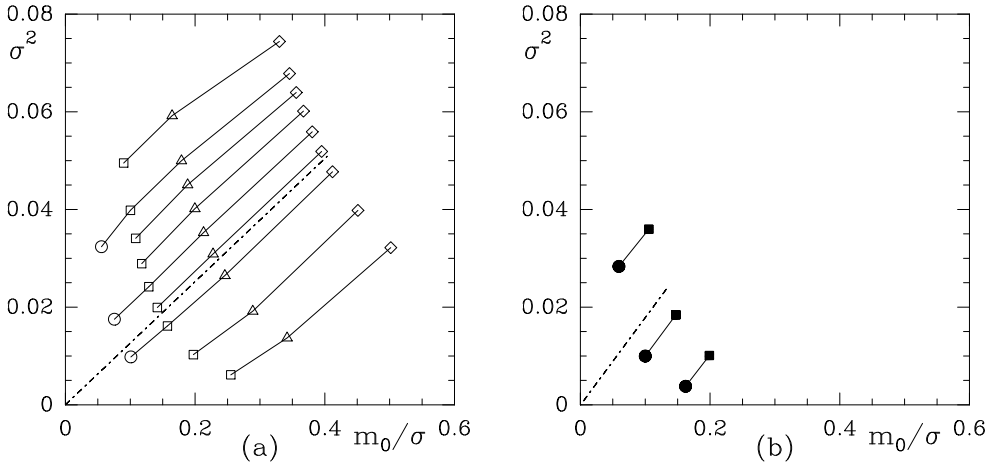
\PSfigure{b}{(14.1,5.5)}{(0,1)}{fig-fisher}{fig-fisher}{Fisher
plots of \MC\ data for the chiral condensate $\sigma$ on
\aaa~$8^3 \times 16$ and \bbb~$12^4$ lattices.}

%------------------------------------------------------------------------
\begin{table}[b]
\begin{center}
\begin{tabular}{|l|l|lll|}
\hline
\multicolumn{5}{|c|}{$\CHI$} \\
\hline
\hline
\multicolumn{1}{|c|}{$g_0$} &
\multicolumn{1}{c|}{$m_0$} &
\multicolumn{1}{c}{$8^4$} &
\multicolumn{1}{c}{$12^4$} &
\multicolumn{1}{c|}{$\infty$} \\
\hline
\hline
0.21 & 0.01 & 0.0294(1) & 0.0284(1) & 0.0275(3) \\
\hline
0.23 & 0.01 & 0.0419(3) & 0.0399(1) & 0.0387(5) \\
\hline
0.25 & 0.01 & 0.0632(8) & 0.0616(4) & 0.0608(11) \\
     & 0.02 & 0.1044(10) & 0.1005(3) & 0.0992(8) \\
\hline
0.27 & 0.01 & 0.1039(12) & 0.0999(10) & 0.0986(18) \\
     & 0.02 & 0.1421(13) & 0.1358(5) & 0.1343(10) \\
\hline
0.28 & 0.01 & 0.1283(11) & 0.1242(9) & 0.1230(15) \\
\hline
0.30 & 0.01 & 0.1675(20) & 0.1683(6) & 0.1685(10) \\
     & 0.02 & 0.1929(12) & 0.1896(4) & 0.1892(6) \\
\hline
0.32 & 0.01 & 0.2021(11) & 0.2042(7) & 0.2045(9) \\
\hline
\end{tabular}
\end{center}
\CAPTION{tab-CHI-11}{\MC\ results for the chiral condensates on lattices with
$L_s:L_t = 1:1$. The third column gives the extrapolation to infinite volume
using \Eq{eq-chi-fs}.}
\end{table}
%------------------------------------------------------------------------
\begin{table}[t]
\begin{center}
\begin{tabular}{|l|l|lll|}
\hline
\multicolumn{5}{|c|}{$\CHI$} \\
\hline
\hline
\multicolumn{1}{|c|}{$g_0$} &
\multicolumn{1}{c|}{$m_0$} &
\multicolumn{1}{c}{$6^3 \times 12$} &
\multicolumn{1}{c}{$8^3 \times 16$} &
\multicolumn{1}{c|}{$\infty$} \\
\hline
\hline
0.21 & 0.02 & 0.1100(7) & 0.0784(2) & 0.0559(18) \\
     & 0.04 & 0.1458(8) & 0.1170(4) & 0.0963(19) \\
     & 0.09 & 0.1995(7) & 0.1794(2) & 0.1688(9) \\
\hline
0.23 & 0.02 & 0.1354(10) & 0.1014(3) & 0.0765(24) \\
     & 0.04 & 0.1685(5) & 0.1385(4) & 0.1182(17) \\
     & 0.09 & 0.2170(3) & 0.1995(3) & 0.1915(7) \\
\hline
0.25 & 0.01 & 0.1358(12) & 0.0991(7) & 0.0720(43) \\
     & 0.02 & 0.1604(8) & 0.1271(5) & 0.1034(25) \\
     & 0.04 & 0.1890(6) & 0.1627(5) & 0.1471(17) \\
     & 0.09 & 0.2339(6) & 0.2185(3) & 0.2124(8) \\
\hline
0.26 & 0.02 & 0.1721(7) & 0.1412(6) & 0.1205(27) \\
     & 0.04 & 0.2014(12) & 0.1757(4) & 0.1618(18) \\
     & 0.09 & 0.2426(6) & 0.2278(3) & 0.2224(7) \\
\hline
0.27 & 0.01 & 0.1592(13) & 0.1324(8) & 0.1139(38) \\
     & 0.02 & 0.1853(9) & 0.1555(5) & 0.1369(24) \\
     & 0.04 & 0.2096(8) & 0.1878(5) & 0.1767(16) \\
     & 0.09 & 0.2483(9) & 0.2364(3) & 0.2325(8) \\
\hline
0.28 & 0.02 & 0.1962(9) & 0.1700(5) & 0.1542(24) \\
     & 0.04 & 0.2197(9) & 0.2004(4) & 0.1915(14) \\
     & 0.09 & 0.2560(6) & 0.2452(3) & 0.2419(7) \\
\hline
0.29 & 0.02 & 0.2068(12) & 0.1846(7) & 0.1730(24) \\
     & 0.04 & 0.2305(8) & 0.2123(3) & 0.2048(10) \\
     & 0.09 & 0.2621(5) & 0.2529(4) & 0.2503(8) \\
\hline
0.30 & 0.01 & 0.1969(14) & 0.1799(7) & 0.1703(30) \\
     & 0.02 & 0.2169(9) & 0.1996(6) & 0.1919(18) \\
     & 0.04 & 0.2383(9) & 0.2235(5) & 0.2180(13) \\
     & 0.09 & 0.2698(6) & 0.2605(3) & 0.2580(6) \\
\hline
0.32 & 0.02 & 0.2356(10) & 0.2225(6) & 0.2172(17) \\
     & 0.04 & 0.2531(9) & 0.2433(5) & 0.2404(11) \\
     & 0.09 & 0.2798(6) & 0.2728(4) & 0.2711(7) \\
\hline
\end{tabular}
\end{center}
\CAPTION{tab-CHI-12}{\MC\ results for the chiral condensates on lattices with
$L_s:L_t = 1:2$. The third column gives the extrapolation to infinite volume
using \Eq{eq-chi-fs}.}
\end{table}
%------------------------------------------------------------------------

\hfig{t}{fig-chi-comp}{Comparison
of \MC\ data ($8^3 \times 16$ lattice, $m_0 = 0.02$) for
the chiral condensate with results from the gap equation (dotted line),
the order $g_0$ (dashed line) and the order $g_0^2$ (solid line) \SDeqs.}

In \Fig{fig-chi-comp} we compare our \MC\ data with results from
the gap equation and our two \SD\ approximations.
(The chiral condensate is the value of the fermion propagator at zero
 distance:
$\sigma = V^{-1} \sum_x G^{\rm full\ }{}^x_x$.)
There is a large
discrepancy between the numerical data and the gap equation. However the
agreement improves considerably as higher orders are included leading to
a maximum discrepancy of about 10\% with our \MC\ data. The
equations also reproduce the large finite size effects.

Applying the \SDeqs\ for infinite volume and zero bare mass reveals that there
is a chiral phase transition and not just a crossover. In the gap equation this
occurs
at
\begin{equation}
    g_c = \frac{1}{8y} \approx 0.201700 \,,
 \end{equation}
 where
\begin{equation}
  y = \int_0^{2\pi} \frac{d^4 k}{(2\pi)^4}\
   \frac{1}{\sum_\mu\sin^2k_\mu}
  = \int_0^{\infty} d \alpha \, e^{-2\alpha} I_0^4(\alpha/2)
   \approx 0.619734 \,,
 \end{equation}
and in the order
$g_0$ case at
\begin{equation}
   g_c = \frac{32y}{(16y-1)^2} \approx 0.249483 \,.
\end{equation}
 In the order $g_0^2$ case we can give no formula but the
 number is $g_c \approx 0.263$.

The semi-quantitative success of these equations motivated us to make a
phenomenological fit to the \MC\ data based on a modified gap equation
\begin{equation}
 \sigma = \frac{c_5}{V} \sum_k \frac{M}{M^2 + \sum_\mu \sin^2 k_\mu},
 \label{eq-pheno}
\end{equation}
where $\sigma$ is the chiral condensate and
\begin{equation}
 M = (c_1 + c_2 g_0 + c_3 g_0^2) \sigma + c_4 m_0 \,.
\end{equation}
The values of the parameters were found to be:
$c_1 = 0.149(1)$, $c_2 = 6.37(1)$, $c_3 = -3.07(1)$,
$c_4 = 0.938(1)$ and $c_5 = 0.959(1)$.
The results of this fit are plotted in \Fig{fig-chi-fit} which includes
data from different lattice sizes showing that the fit equation also
accommodates quite nicely the \FS\ effects.
 Extrapolating the
phenomenological equation to infinite volume and zero mass gives
 $g_c \approx 0.278$. Comparing this value with the above
results gives a shift of
about 0.015 relative to the order $g_0^2$ \SDeqs.

\hfig{t}{fig-chi-fit}{\MC\
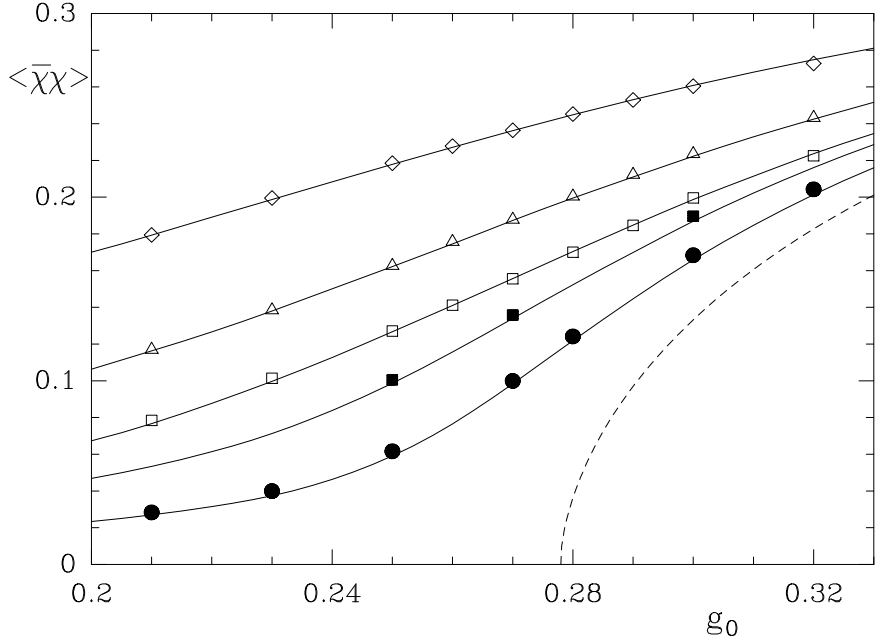
data for the chiral condensate and curves resulting from the
phenomenological fit \Eq{eq-pheno}. The dashed line indicates the
extrapolation to infinite volume and zero bare mass. Symbol shape denotes bare
mass (diamonds $0.09$, triangles $0.04$, squares $0.02$
and circles $0.01$). Colour
denotes lattices size (white $8^3\times16$ and black $12^4$).}

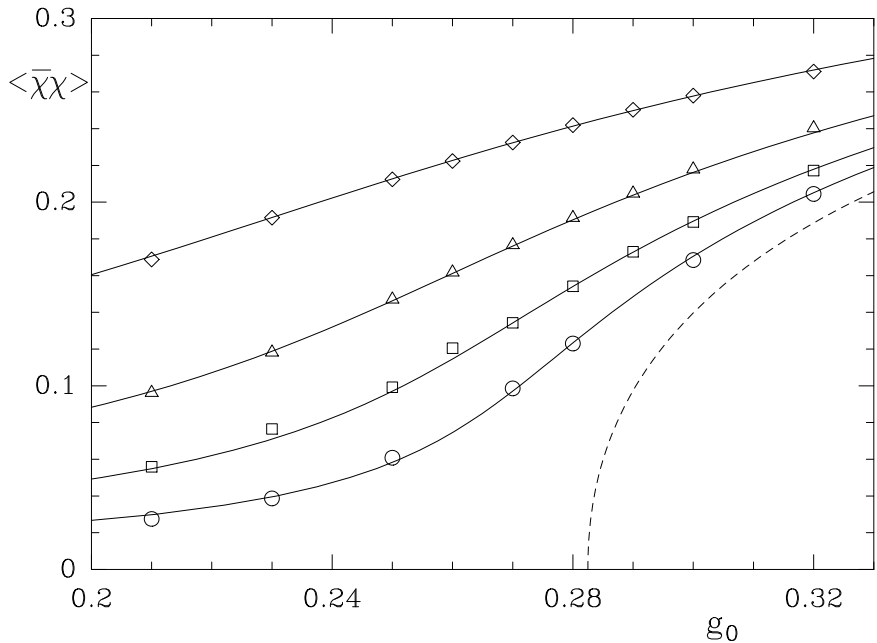
\hfig{t}{fig-chi-mf}{\MC\ data for the chiral condensate extrapolated
 to infinite volume and curves resulting from the mean-field fit
 with logarithmic corrections \Eq{eq-mf}. The dashed line indicates
 the extrapolation to zero bare mass. Symbol shape denotes
 bare mass (diamonds $0.09$, triangles $0.04$,
squares $0.02$ and circles $0.01$).}

The \SDeqs\ as well as the fit equation in the infinite volume limit
lead to the following scaling laws
sufficiently close to the critical point:
\begin{equation}
 \sigma^2 \ln(\sigma_0 /\sigma) \propto g_0 - g_c
 \label{eq-sigma2}
\end{equation}
on the critical line $m_0 = 0$ and
\begin{equation}
 \sigma^3 \ln(\sigma_0 /\sigma)   \propto m_0
 \label{eq-sigma3}
\end{equation}
 at $g_0 = g_c$ ($\sigma_0$ is a constant of order 1).
 This is mean-field behaviour with logarithmic corrections.
Although at this order the power of the logarithm is $+1$
it is possible that the inclusion of higher order
 Feynman diagrams could change this power.

For a more direct determination of the critical exponents and
logarithmic corrections we extrapolate the data to the infinite volume
first and then fit the result with a simple effective potential. The
extrapolation of the chiral condensate to the infinite volume is done
in section 3.3, and the results are stated in tables 1 and 2.
Wherever we have two extrapolations, we take the one coming from the
larger volume. We have performed two fits. In the first fit we assume
an equation of state of the form
\begin{equation}
m_0 = (\delta - \frac{1}{\beta} + 1) \kappa \sigma^{\delta
       - \frac{1}{\beta}} + (\delta + 1) \zeta \sigma^{\delta},
\end{equation}
with
\begin{eqnarray}
\kappa & = & \kappa_1 (g_0 - g_c), \\
\zeta  & = & \zeta_0 + \zeta_1 (g_0 - g_c),
\end{eqnarray}
thus leaving the critical exponents as free parameters. We find
$\kappa_1 = -3.1(1)$, $\zeta_0 = 2.6(2)$,
$\zeta_1 = 8(1)$, $\beta = 0.42(1)$,
$\delta = 3.4(1)$ and $g_c = 0.280(1)$.
The exponents $\beta$ and $\delta$ lie rather close to their
mean field values ($1/2$ and $3$ respectively).
In the second fit we fix $\beta$
and $\delta$ to their mean field values but allow for logarithmic
corrections:
\begin{equation}
m_0 = 2 \kappa \frac{\sigma}{{\rm ln}^p|\sigma^{-1}|} +
      4 \zeta \frac{\sigma^3}{{\rm ln}^q|\sigma^{-1}|}.
\label{eq-mf}
\end{equation}

The result of the fit is $\kappa_1 = -3.6(3)$, $\zeta_0 = 2.1(3)$,
$\zeta_1 = 9(2)$, $p = 0.36(11)$,
$q = 0.84(1)$ and $g_c = 0.282(2)$. The
data and the fit are plotted in \Fig{fig-chi-mf}.
 (The former fit leads to
similar curves.)
It is pleasing to note that all three fits give roughly the same value of
$g_c$.
 It is however possible that the powers $p$ and $q$ found from
 the fit \Eq{eq-mf} have not yet reached their asymptotic values
 in the parameter region we have covered.
%Because the equation of state one derives from \Eq{eq-pheno} is an
%infinite series in $\sigma$ and $\ln \sigma$, the
%asymptotic scaling laws (\ref{eq-sigma2}), (\ref{eq-sigma3})
%may not be effective in the parameter region we
%have covered.

\hfig{b}{fig-banana}{\MC\
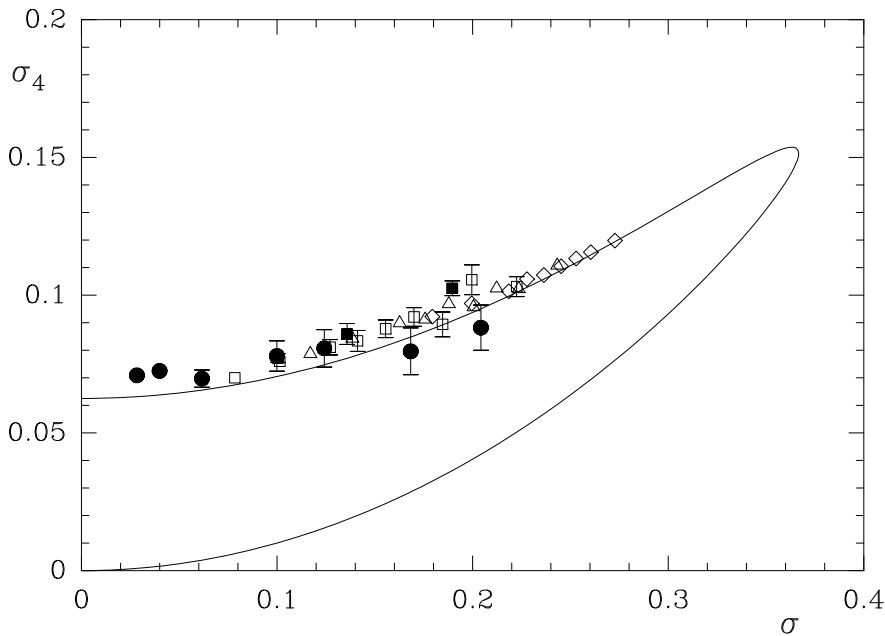
results for the four fermion condensate plotted against the chiral
condensate. The curve results from eqs.~(\ref{banana.1})
and~(\ref{banana.2}). The symbols are the same as in \Fig{fig-chi-fit}.}

To gain a first impression of the short distance behaviour of the model
we have taken our data and in \Fig{fig-banana} plotted the four fermion
condensate,
$\sigma_4 = \mbox{\small${1\over 4}$} \sum_{x,\mu} \langle
              \bar{\chi}(x) \chi(x)
              \bar{\chi}(x+\hat{\mu}) \chi(x+\hat{\mu}) \rangle /V $,
against the chiral condensate. $\sigma_4$ was
also calculated via a stochastic estimator, but now involves two
fermion matrix inversions. (The number of sets of random numbers
used and hence inversions performed was also increased from one for
the chiral condensate to $\sim O(10\%)$ of the number of molecular
dynamics steps in the Hybrid Monte Carlo program.) The results are
given in \Tab{tab-chi4}. The points appear to lie roughly on one curve.
We have compared the Monte Carlo data with the naive result obtained
from free fermion propagators. This gives
\begin{eqnarray}
  \sigma_4(m) &=& {1\over 4}\sum_\nu \left[
                  {1\over V} \sum_p { {\sin^2 p_\nu} \over
                                      {\sum_\lambda \sin^2 p_\lambda + m^2}}
                                     \right]^2
                  + \sigma(m)^2     \nonumber \\
                &\stackrel{V\to\infty}{\to}&
                   {1\over 16}  \left[ 1 - m \sigma(m) \right]^2
                  + \sigma(m)^2
\label{banana.1}
\end{eqnarray}
against
\begin{eqnarray}
  \sigma(m)  &=&  {m \over V} \sum_p {1\over \sum_\lambda \sin^2 p_\lambda
                                          + m^2}  \nonumber \\
                &\stackrel{V\to\infty}{\to}&
                   \int_{-\pi}^{\pi} {d^4 p \over (2\pi)^4}
                     {m \over \sum_\lambda \sin^2 p_\lambda + m^2}\,,
\label{banana.2}
\end{eqnarray}
 where $m$ is to be regarded as a parameter.
The infinite volume result is also plotted in \Fig{fig-banana}. The data lie
close to the upper branch (this corresponds to small $\mR$, while
the lower branch corresponds to large $\mR$). This similarity between
the data and the free fermion formula suggests that there is only
 little interaction between the fermions, the main effect of the
interaction being the generation of a renormalised mass, the determination
of which we shall now turn to.

%------------------------------------------------------------------------
\begin{table}[t]
\begin{center}
\begin{tabular}{|c|ll|lll|}
\hline
\multicolumn{6}{|c|}{$\sigma_4$} \\
\hline
\hline
& \multicolumn{2}{c|}{$12^4$}&\multicolumn{3}{c|}{$8^3\times 16$}\\
\cline{2-6}
$g_0$&$m_0=0.01$ & $m_0=0.02$ & $m_0=0.02$ & $m_0=0.04$ & $m_0=0.09$\\
\hline
\hline
0.21 & 0.071(1) &          & 0.070(3) & 0.079(1) & 0.0922(4) \\
0.23 & 0.072(2) &          & 0.076(3) & 0.084(2) & 0.0970(5) \\
0.25 & 0.070(3) & 0.077(2) & 0.081(3) & 0.090(1) & 0.1014(5) \\
0.26 &          &          & 0.083(4) & 0.091(1) & 0.1058(6) \\
0.27 & 0.078(6) & 0.086(4) & 0.088(3) & 0.097(2) & 0.1073(5) \\
0.28 & 0.081(7) &          & 0.092(3) & 0.096(1) & 0.1105(4) \\
0.29 &          &          & 0.089(4) & 0.102(1) & 0.1133(6) \\
0.30 & 0.080(8) & 0.102(3) & 0.106(5) & 0.102(1) & 0.1156(5) \\
0.32 & 0.088(8) &          & 0.103(4) & 0.111(3) & 0.1198(6) \\
\hline
\end{tabular}
\end{center}
\CAPTION{tab-chi4}{\MC\ results for the four fermion condensate $\sigma_4$.}
\end{table}
%------------------------------------------------------------------------

\subsection{Fermion Propagator}

%------------------------------------------------------------------------
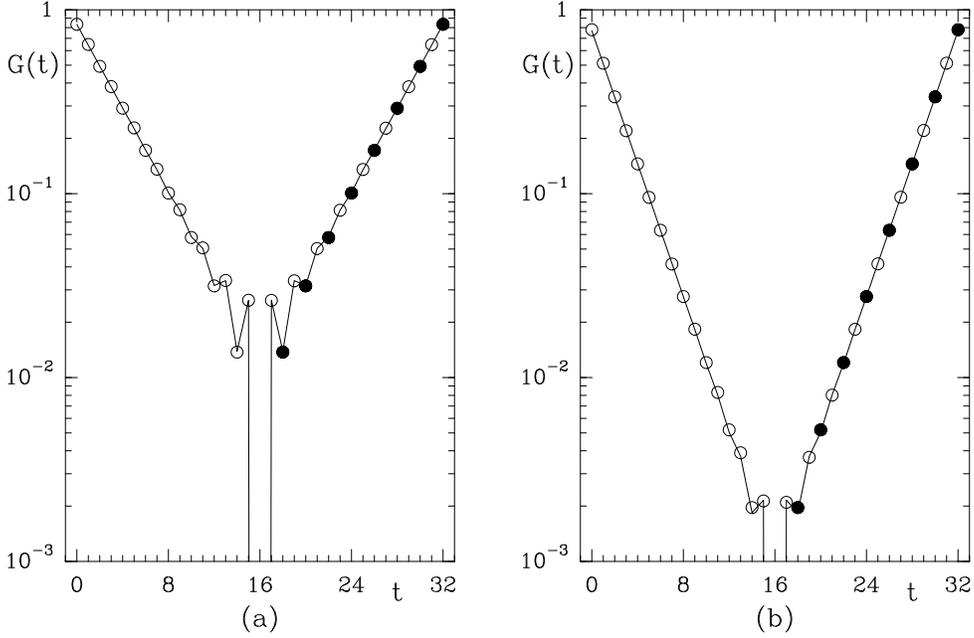
\hfig{t}{fig-prop-mr}{Fermion
propagators from \MC\ calculations on an $8^3 \times 32$ lattice at
$m_0 =0.04$, $g_0 = 0.23$ \aaa\ and $g_0 = 0.28$ \bbb. Open (solid) symbols
represent positive (negative) values. The lines connect values from fits using
\Eq{eq-G-fit}. The resulting fermion masses are \aaa\ $\mu_R = 0.262(1) $ and
 \bbb\ $\mu_R = 0.417(1) $.}

%------------------------------------------------------------------------
\begin{table}[b]
\begin{center}
\begin{tabular}{|c|ll|lll|}
\hline
\multicolumn{6}{|c|}{$Z_2$} \\
\hline
\hline
& \multicolumn{2}{c|}{$12^4$}&\multicolumn{3}{c|}{$8^3\times 16$}\\
\cline{2-6}
$g_0$&$m_0=0.01$ & $m_0=0.02$ & $m_0=0.02$ & $m_0=0.04$ & $m_0=0.09$\\
\hline
\hline
0.21 & 0.872(1) &          & 0.865(1) & 0.872(1) & 0.887(1) \\
0.23 & 0.857(1) &          & 0.858(2) & 0.867(2) & 0.880(1) \\
0.25 & 0.834(1) & 0.848(1) & 0.842(2) & 0.858(2) & 0.871(2) \\
0.26 &          &          &          & 0.854(2) & 0.865(2) \\
0.27 & 0.825(2) & 0.830(2) & 0.835(3) & 0.847(2) & 0.866(2) \\
0.28 & 0.817(2) &          & 0.836(3) & 0.845(2) & 0.865(2) \\
0.29 &          &          & 0.831(3) & 0.845(1) & 0.860(2) \\
0.30 & 0.816(3) & 0.832(2) & 0.832(3) & 0.848(4) & 0.862(2) \\
0.32 & 0.827(3) &          & 0.825(4) & 0.844(3) & 0.853(3) \\
\hline
\end{tabular}
\end{center}
\CAPTION{tab-Z2}{\MC\ results for the wave-function renormalisation
constant $Z_2$.}
\end{table}

\noindent To determine the fermion mass
 the fermion propagator
\begin{equation}
 G(t) = \sum_{\vec{x} \atop x_1, x_2, x_3\, {\rm even}}
    \langle \chi(\vec{x}, t) \bar{\chi}(\vec{0}, 0) \rangle
\end{equation}
is fitted by
\begin{equation}
 G(t) = \frac{Z_2}{\left(1 + e^{-\mR L_t}\right) \cosh \mR}
 \left( e^{-\mR t} - (-1)^t e^{-\mR (L_t - t)} \right).
 \label{eq-G-fit}
\end{equation}
The parameter $\mR$ is the renormalised fermion mass
and $Z_2$ is the wave-function renormalisation
constant. The results of the fits are
given in tables~\ref{tab-Z2}, \ref{tab-mR-11} and~\ref{tab-mR-12}.
 In \Fig{fig-prop-mr} we show two propagators on $8^3 \times 32$
 lattices which were simulated to check the masses obtained on lattices
 with smaller time extent.
% Two representative propagators are shown in
%\Fig{fig-prop-mr}.
 We see that their behaviour is very well described by
a single exponential.

In \Fig{fig-chi-mr} we plot $\CHI$ against $\mR$.
In this plot we have also shown the predictions of
the gap equation and of the order $g_0^2$ \SDeqs.
We see that the data lie in a narrow band. The
theoretical curves also display a small width because they
still have weak dependences on bare mass and lattice size.
The small spread tells us that the
form of the fermion propagator is hardly changed by the interactions.

\ifig{t}{fig-chi-mr}{\MC\
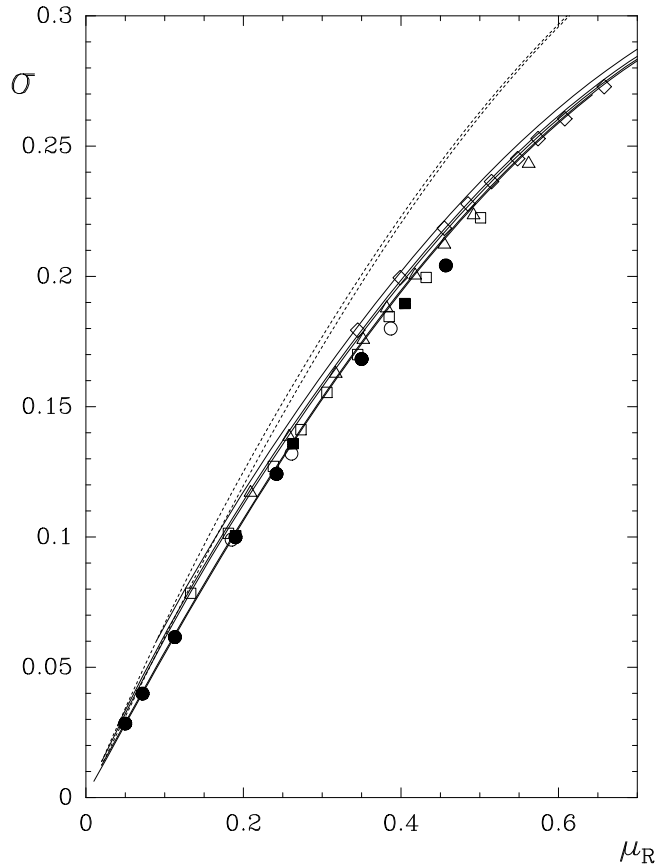
results for the chiral condensate plotted versus the renormalised
fermion mass. The dotted lines are results from the gap equation. The solid
lines are results from the order $g_0^2$ \SDeqs. The symbols are the same as in
\Fig{fig-chi-fit}.}

\subsection{Finite Size Effects} \label{sect-fs-chi}

    Comparing results found on  lattices with different volumes (see
tables~\ref{tab-CHI-11} and~\ref{tab-CHI-12}) makes it clear that we have large
\FS\ effects  in our data, and that it is important to study these in order  to
see the picture for an infinite lattice.  An advantage   we have over the usual
situation in a \MC\   calculation is that we can use the \SDeqs\   to calculate
results on large or even on infinite lattices,   and so see how the
thermodynamic limit is approached.

%------------------------------------------------------------------------
\begin{table}[t]
\begin{center}
\begin{tabular}{|l|l|lll|}
\hline
\multicolumn{5}{|c|}{$\mR$} \\
\hline
\hline
\multicolumn{1}{|c|}{$g_0$} &
\multicolumn{1}{c|}{$m_0$} &
\multicolumn{1}{c}{$8^4$} &
\multicolumn{1}{c}{$12^4$} &
\multicolumn{1}{c|}{$\infty$} \\
\hline
\hline
0.21 & 0.01 & 0.051(0) & 0.050(0) & 0.050(1) \\
\hline
0.23 & 0.01 & 0.075(1) & 0.072(0) & 0.070(1) \\
\hline
0.25 & 0.01 & 0.116(1) & 0.113(0) & 0.112(2) \\
     & 0.02 & 0.196(2) & 0.190(0) & 0.188(1) \\
\hline
0.27 & 0.01 & 0.202(3) & 0.190(1) & 0.185(3) \\
     & 0.02 & 0.274(2) & 0.263(1) & 0.260(2) \\
\hline
0.28 & 0.01 & 0.258(3) & 0.242(1) & 0.237(2) \\
\hline
0.30 & 0.01 & 0.352(4) & 0.350(1) & 0.350(2) \\
     & 0.02 & 0.412(3) & 0.405(1) & 0.404(2) \\
\hline
0.32 & 0.01 & 0.460(6) & 0.457(2) & 0.456(2) \\
\hline
\end{tabular}
\end{center}
\CAPTION{tab-mR-11}{\MC\ results for the renormalised fermion mass on lattices
with $L_s:L_t = 1:1$. The third column gives the extrapolation to infinite
volume using \Eq{eq-mR-fs}.}
\end{table}

    To keep the calculations reasonably simple we will  look at the lattice
size dependence of the solutions
of the order $g_0$ \SDeqs~(\ref{OGfermi}), but we expect
that  the form of the formulae found will remain true in general.

    The equations (\ref{OGfermi}) always involve the sum over allowed   momenta
of some function $a(k)$ of the momentum.   Let us call such a sum $U_a$.     We
need to compare this sum over momenta   with the integral over all momenta
which gives the infinite volume limit. We do this  by  using the Poisson
resummation formula to rewrite   our sum as an integral
 \begin{eqnarray}
    U_a & = & \frac{1}{V} \sum_k a(k) \nonumber \\
        & = & \int d^4k' \frac{1}{V} \sum_k a(k') \delta(k' - k) \,.
 \end{eqnarray}
The sum of $\delta$ functions can be expressed as a Fourier series, leading to
 \begin{equation}
  U_a = \sum_n (-1)^{n_t} A(\vec{n} L_s, n_t L_t) \,,
 \end{equation}
where $A$ is the Fourier transform of $a$ defined by
\begin{equation}
  A(x) = \int\frac{d^4k}{(2\pi)^4} e^{ik \cdot x} a(k)
\end{equation}
(the $(-1)^{n_t}$ arises because of the antiperiodic boundary conditions in
time).
   The functions that we consider will always have   Fourier transforms which
decrease exponentially at   large distance.  Therefore   in an infinite volume
only $A(0)$ survives,   while on a large lattice the finite size effects   will
be dominated by terms where one of   the $n_\mu$ is $\pm 1$.  Physically
speaking these   represent the contributions of particles   which have
travelled ``once around the lattice". Terms   where two of the $n_\mu$'s are
non-zero, or one of the $n_\mu$'s is   larger than 1 die away as exponentials
with higher exponent   than the leading correction and very quickly become
negligible.

%------------------------------------------------------------------------
\begin{table}[t]
\begin{center}
\begin{tabular}{|l|l|lll|}
\hline
\multicolumn{5}{|c|}{$\mR$} \\
\hline
\hline
\multicolumn{1}{|c|}{$g_0$} &
\multicolumn{1}{c|}{$m_0$} &
\multicolumn{1}{c}{$6^3 \times 12$} &
\multicolumn{1}{c}{$8^3 \times 16$} &
\multicolumn{1}{c|}{$\infty$} \\
\hline
\hline
0.21 & 0.02 & 0.183(1) & 0.133(0) & 0.098(3) \\
     & 0.04 & 0.250(1) & 0.209(0) & 0.179(2) \\
     & 0.09 & 0.377(1) & 0.345(0) & 0.328(1) \\
\hline
0.23 & 0.02 & 0.240(2) & 0.181(0) & 0.138(4) \\
     & 0.04 & 0.311(1) & 0.258(0) & 0.222(3) \\
     & 0.09 & 0.430(1) & 0.399(0) & 0.385(1) \\
\hline
0.25 & 0.01 & 0.268(3) & 0.185(1) & 0.123(8) \\
     & 0.02 & 0.305(2) & 0.238(1) & 0.191(4) \\
     & 0.04 & 0.364(2) & 0.317(1) & 0.289(3) \\
     & 0.09 & 0.483(1) & 0.455(1) & 0.445(2) \\
\hline
0.26 & 0.02 & 0.334(2) & 0.273(1) & 0.232(6) \\
     & 0.04 & 0.397(2) & 0.352(1) & 0.328(3) \\
     & 0.09 & 0.512(1) & 0.485(1) & 0.475(2) \\
\hline
0.27 & 0.01 & 0.329(3) & 0.261(1) & 0.213(8) \\
     & 0.02 & 0.365(2) & 0.306(1) & 0.269(5) \\
     & 0.04 & 0.430(2) & 0.382(1) & 0.358(3) \\
     & 0.09 & 0.538(1) & 0.515(1) & 0.508(2) \\
\hline
0.28 & 0.02 & 0.416(3) & 0.345(1) & 0.303(6) \\
     & 0.04 & 0.462(2) & 0.418(1) & 0.398(3) \\
     & 0.09 & 0.572(2) & 0.548(1) & 0.541(2) \\
\hline
0.29 & 0.02 & 0.442(3) & 0.385(1) & 0.355(6) \\
     & 0.04 & 0.492(2) & 0.455(1) & 0.440(2) \\
     & 0.09 & 0.593(2) & 0.574(1) & 0.569(2) \\
\hline
0.30 & 0.01 & 0.463(5) & 0.387(2) & 0.343(10) \\
     & 0.02 & 0.476(3) & 0.432(1) & 0.412(5) \\
     & 0.04 & 0.534(3) & 0.492(2) & 0.477(4) \\
     & 0.09 & 0.630(2) & 0.608(1) & 0.602(2) \\
\hline
0.32 & 0.02 & 0.563(5) & 0.501(2) & 0.475(7) \\
     & 0.04 & 0.585(3) & 0.562(1) & 0.555(3) \\
     & 0.09 & 0.681(2) & 0.658(1) & 0.653(2) \\
\hline
\end{tabular}
\end{center}
\CAPTION{tab-mR-12}{\MC\ results for the renormalised fermion mass on lattices
with $L_s:L_t = 1:2$. The third column gives the extrapolation to infinite
volume using \Eq{eq-mR-fs}.}
\end{table}
%------------------------------------------------------------------------

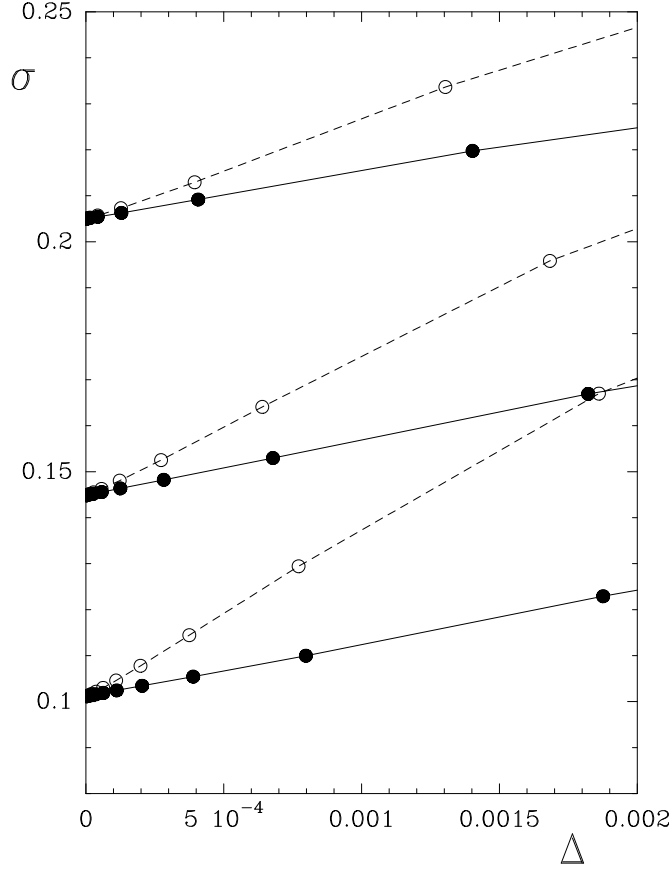
\ifig{t}{fig-chi-fs-sd}{Finite
size scaling of the chiral condensate. Plotted are \SD\ results on
lattices with $L_s \geq 6$;
$L_s:L_t = 1:1$ (solid symbols) and $L_s:L_t = 1:2$
(open symbols). The parameters are $m_0 = 0.02$ and $g_0 = 0.23, 0.25, 0.28$.}

To see what the \FS\ corrections to the fermion   propagator are we need to
know the large distance   behaviour of the Fourier transforms of
  \begin{equation}
  \frac{N}{ N^2 +\sum_\nu F_\nu^2 \sin^2 k_\nu }
  \end{equation}
and
  \begin{equation}
      \frac{ F_\mu \sin^2 k_\mu}
     {N^2 +\sum_\nu F_\nu^2 \sin^2 k_\nu} \,, \label{eq-64}
  \end{equation}
 i.~e., the large distance behaviour of the fermion propagator.
   These limits can be found by the
  saddle point approximation.
For definiteness we look at the case $n_1 = 1$.
First integrate
over $k_1$ which can be done exactly by looking
at residues at the poles of the integrand:

\clearpage

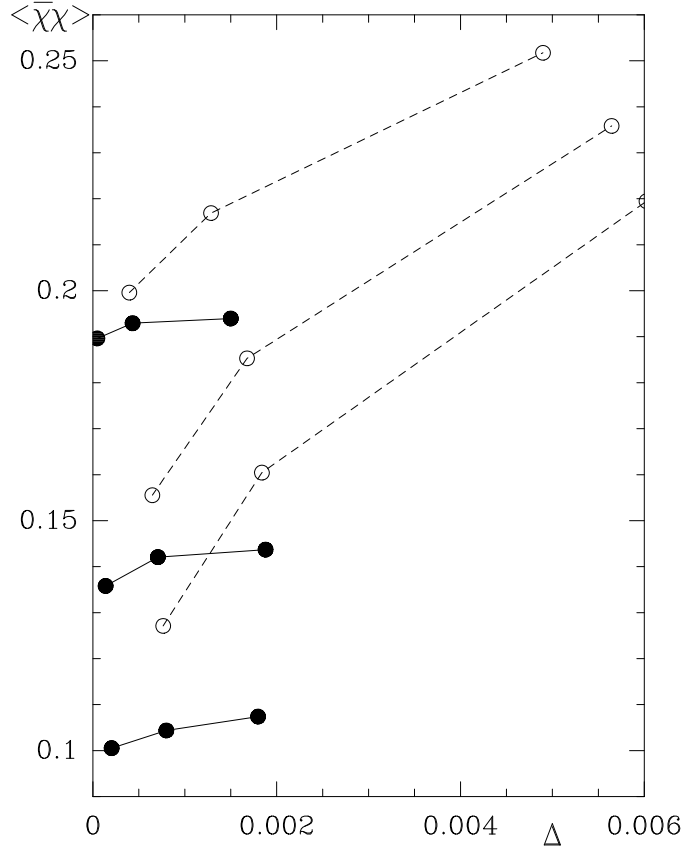
\ifig{t}{fig-chi-fs-mc}{Finite
size scaling of the chiral condensate (\MC\ results).
The lattice sizes are $6^4$, $8^4$ and $12^4$ $(\bullet)$ and
$4^3\times8$, $6^3\times12$ and $8^3\times16$ $(\circ)$.
The parameters are
$m_0 = 0.02$ and $g_0 = 0.25, 0.27, 0.30$.}

 \begin{eqnarray}
 & &  \int \frac {d^4 k}{(2 \pi)^4} \exp( i k_1 L_s )
  \frac {N}{ N^2 +\sum_\nu F_\nu^2 \sin^2 k_\nu } \nonumber \\
 & \stackrel{L_s\, {\rm even}}{=} &
  \int \frac {d k_2 d k_3 d k_4}{(2 \pi)^3}
  \frac{1}{\sqrt { N^2 +\sum_{\nu \neq 1} F_\nu^2 \sin^2 k_\nu }}
  \frac{N}{\sqrt {N^2 + F_1^2 +\sum_{\nu \neq 1} F_\nu^2 \sin^2 k_\nu }}
 \nonumber \\
 & \times &  \exp \left[ - L_s  \sinh^{-1}
    \left( \frac {\sqrt{ N^2 +\sum_{\nu \neq 1} F_\nu^2 \sin^2 k_\nu} }
  {F_1} \right)  \right] \,.
\end{eqnarray}
For large $L_s$ this integral
  is dominated by $\sin k_\nu \approx 0$ and is approximately
  Gaussian:
\begin{eqnarray}
 &\approx&
  \left(\frac {2 N}{\pi L_s} \right)^{\frac{3}{2}}
   \frac {(N^2 + F_1^2)^{1/4} }{F_2 F_3 F_4}
  \exp \left[ -L_s  \sinh^{-1} (N/F_1) \right] \nonumber \\
  &\propto &\left(\mR /L_s \right)^{\frac{3}{2}} \exp ( -\mR L_s ) \,.
  \end{eqnarray}
  (In the final expression $\mR \ll 1$ has been taken.)   The change in the sum
for \Eq{eq-64} also has   the same $L_s$ dependence. The propagators in the 2-
and 3- direction have the same form. In the 4-direction $L_s$ goes to $L_t$.

 Naturally the eigenvalues of the transfer matrix are independent of   the
temporal extension of the lattice, nevertheless the rate at   which correlation
functions decay can depend on $L_t$, and it is this   decay rate which is used
to measure $\mu_R$ at finite $L_t$.   The infinite volume limit turns out to be
independent of $L_t/L_s$.

   The change in the solution of (\ref{OGfermi}) will be proportional  to the
difference between the   finite volume sum and the infinite volume integral
when this difference is itself small. This leads to the   expectation that for
$L_t \geq L_s$
  \begin{equation}
   \mR - \mR(\infty) \propto  g_0 (\mR /L_s)^{\frac{3}{2}} \exp ( -\mR L_s )
  \label{eq-mR-fs}
  \end{equation}
   and
  \begin{equation}
   \CHI -\CHI(\infty) \propto (\mR /L_s)^{\frac{3}{2}} \exp ( -\mR L_s )
       \equiv \Delta \,.
  \label{eq-chi-fs}
  \end{equation}
 \noindent  These equations are to be understood as giving the form
  of the $L_s$ dependence at fixed bare parameters for
  large $L_s$. The constant of
  proportionality depends on the values
  of the bare parameters.

      The above predictions have been tested by plotting
   $\CHI$ or $\mR$ against the right-hand side of (\ref{eq-mR-fs}),
   (\ref{eq-chi-fs}) respectively. Data for a given lattice
   shape should fall on a straight line.
  In \Fig{fig-chi-fs-sd} we show the \SD\ results for the chiral condensate and
see that the asymptotic formula (\ref{eq-chi-fs})
holds for $L_s \geq 6$. A similar
conclusion holds for the \MC\ results as \Fig{fig-chi-fs-mc} reveals.
  We gain added faith in this extrapolation
  to infinite volume from the fact that lattices with $L_t = L_s$
  and with $L_t = 2 L_s$ both extrapolate to the same value
  at infinite volumes.
The extrapolated values for $\CHI$ and $\mR$ are given in tables
\ref{tab-CHI-11}, \ref{tab-CHI-12}, \ref{tab-mR-11} and \ref{tab-mR-12}.
It is pleasing to note that the results from the $12^4$ lattice already
lie very close to the infinite volume results.

The above formulae are only valid in a domain where
$\mbox{\it any mass} \times L_s$ is
large. Near the critical line $m_\pi \times L_s$ is very small and the \FS\
formulae of~\cite{Leutwyler} should apply. Near the critical point
$\mbox{\it all masses} \times L_s$
are very small and the exponential corrections turn into power law
corrections.

\clearpage

\section{Fermion-Antifermion Composite States}

We now turn to the determination of the energies
of fermion-antifermion composite states (``mesons'').

\subsection{\MC\ Method}

%------------------------------------------------------------------------
\begin{table}[b]
\begin{center}
\begin{tabular}{||c|c||c|c||c|c||}
\hline
$i$ & $s_{i,k}(\vec{x},t)$      & $J^{PC}$   & name    & $J^{PC}$   & name \\
\hline
\hline
\multicolumn{6}{||c||}{local operators} \\
\hline
1  & $(-1)^t$                   & $0_a^{-+}$ & $\pi'$  & $0_s^{++}$ & $\sigma$\\
\hline
2  & $(-1)^{x_1+x_2+x_3+t}$     & $0_a^{-+}$ & $\pi$   & $0_a^{+-}$ & $-$ \\
\hline
3  & $(-1)^{x_k+t}$             & $1_a^{--}$ & $\rho'$ & $1_a^{++}$ & $a$ \\
\hline
4  & $(-1)^{x_1+x_2+x_3-x_k+t}$ & $1_a^{--}$ & $\rho$  & $1_a^{+-}$ & $b$ \\
\hline
\hline
\multicolumn{6}{||c||}{one-bond operator} \\
\hline
 & $\eta_k (\vec{x}, t)$        & $1_s^{--}$ &$\omega$ & $1_a^{+-}$ & $b'$ \\
\hline
\hline
\multicolumn{2}{||l||}{terms in \Eq{eq-fit}:} &
\multicolumn{2}{c||}{$E^{+}$} & \multicolumn{2}{c||}{$E^{-}$} \\
\hline
\end{tabular}
\end{center}
\CAPTION{tab-names}{Sign factors, corresponding continuum quantum number
assignments and names of the particles that are used in the text. The last line
gives the connection to the fit function, \Eq{eq-fit}.}
\end{table}
%------------------------------------------------------------------------

\noindent We calculated
the correlation functions of bilinear operators of definite lattice
symmetry which are well known from QCD spectroscopy (see,
e.~g.,~\cite{Golterman,Alt92}).
We looked at four local operators and a
one bond operator. For results on other one bond operators see~\cite{thesis}.
The propagators for the local-local correlation functions
have been calculated with 48 sources.
Additionally we considered four wall operators~\cite{wall}.
The local operators ${\cal O}^{(0)}$ are given by
\begin{equation}
 {\cal O}_{i,k}^{(0)}(t) =
    \sum_{\vec{x}} s_{i,k}(\vec{x},t) \bar{\chi}(\vec{x},t) \chi(\vec{x},t) \,,
\end{equation}
where the $s$ factors and the corresponding continuum quantum number
assignments are given in the \Tab{tab-names}.
The notation for the continuum quantum numbers is the standard $J^{PC}$.
The subscript is the $SU(4)$ flavour representation ($a$ for adjoint, $s$ for
singlet) under the assumption that the $SU(4)$ flavour symmetry is restored.
Our fifth operator is
\begin{equation}
 {\cal O}_{1,k}^{(1)}(t) = \frac{1}{2}
   \sum_{\vec{x}} \eta_k(\vec{x},t) \bar{\chi}(\vec{x},t)
   \left[ \chi(\vec{x} + \hat{k}, t) + \chi(\vec{x} - \hat{k}, t)
   \right] \,.
\end{equation}
For the wall operators we take
\begin{equation}
 {\cal O}_{i,k}^{\rm (w)}(t) = \left(\frac{2}{L_s}\right)^3
   \sum_{\vec{x}, \vec{e}} s_{i,k}(\vec{x}, t)
   \bar{\chi}(\vec{x} + \vec{e}, t) \chi(\vec{x}, t) \,.
\end{equation}
The $\vec{e}$ sum extends over the vectors with all coordinates even.
Correlation functions are
\begin{eqnarray}
 C_i^{(l)}(t) & = &
  \langle {\cal O}_{i,k}^{(l)}(t) {\cal O}_{i,k}^{(l)}(0) \rangle
  - \langle {\cal O}_{i,k}^{(l)}(t) \rangle
    \langle {\cal O}_{i,k}^{(l)}(0) \rangle, \; l = 0,1, \\
 C_i^{\rm (w)}(t) & = &
  \langle {\cal O}_{i,k}^{(0)}(t) {\cal O}_{i,k}^{\rm (w)}(0) \rangle
  - \langle {\cal O}_{i,k}^{(0)}(t) \rangle
    \langle {\cal O}_{i,k}^{\rm (w)}(0) \rangle \,.
\end{eqnarray}
They are independent of $k$. Whenever several values of $k$ are possible we
averaged over them.

These correlation functions are the sum of a fermion line connected and
a fermion line disconnected part (annihilation part). Pictorially this
is represented as
%
%vvvvvvvvvvvvvvvvvvvvvvvvvvvvvvvvvvvvvvvvvvvvvvvvvvvvvvvvvvvvvvvvvvvvvvvvvvvvvvv
%
\begin{equation}  % Begin of pictorial equation.
\newcommand{\CONN}{\mbox{\raisebox{0.618ex}[0.5cm][0.5cm]{\unitlength 1cm
\begin{picture}(2,1)
\put(0,0){\circle*{0.25}}
\put(2,0){\circle*{0.25}}
\thicklines
\put(1.1,.5){\vector(1,0){0}}
\put(0.9,-.5){\vector(-1,0){0}}
\thinlines
\bezier{200}(0,0)(1,-1)(2,0)
\bezier{200}(0,0)(1,1)(2,0)
\end{picture}}}}
\newcommand{\DISL}{\mbox{\raisebox{0.618ex}[0.5cm][0.5cm]{\unitlength 1cm
\begin{picture}(0.8,0.8)
\put(0.05,0){\circle*{0.25}}
\put(0.4,0){\circle{0.7}}
\thicklines
\put(0.75,0.1){\vector(0,1){0}}
\end{picture}}}}
\newcommand{\DISR}{\mbox{\raisebox{0.618ex}[0.5cm][0.5cm]{\unitlength 1cm
\begin{picture}(0.8,0.8)
\put(0.75,0){\circle*{0.25}}
\put(0.4,0){\circle{0.7}}
\thicklines
\put(0.05,-0.1){\vector(0,-1){0}}
\end{picture}}}}
 C = \left\langle\, \CONN\, \right\rangle
   + \left\langle \DISL\,\,\,\DISR \right\rangle
   - \left\langle \DISL \right\rangle  \left\langle \DISR \right\rangle,
\label{eq-picture}
\end{equation} % End of pictorial equation.
%
%^^^^^^^^^^^^^^^^^^^^^^^^^^^^^^^^^^^^^^^^^^^^^^^^^^^^^^^^^^^^^^^^^^^^^^^^^^^^^^^
%
where the brackets indicate averaging over the auxiliary field. Usually
one measures only the fermion line connected part
which is the first term in (\ref{eq-picture}). The second term, which we call
the annihilation
part, is computationally very demanding because the noise problem is more
severe. However we have also attempted to measure the
annihilation part with a stochastic estimator. The last term only
contributes for particles with vacuum quantum numbers.

%------------------------------------------------------------------------
\PSfigure{p}{(14.1,19.0)}{(0,0.7)}{fig-prop-mc}{fig-prop-mc}{%
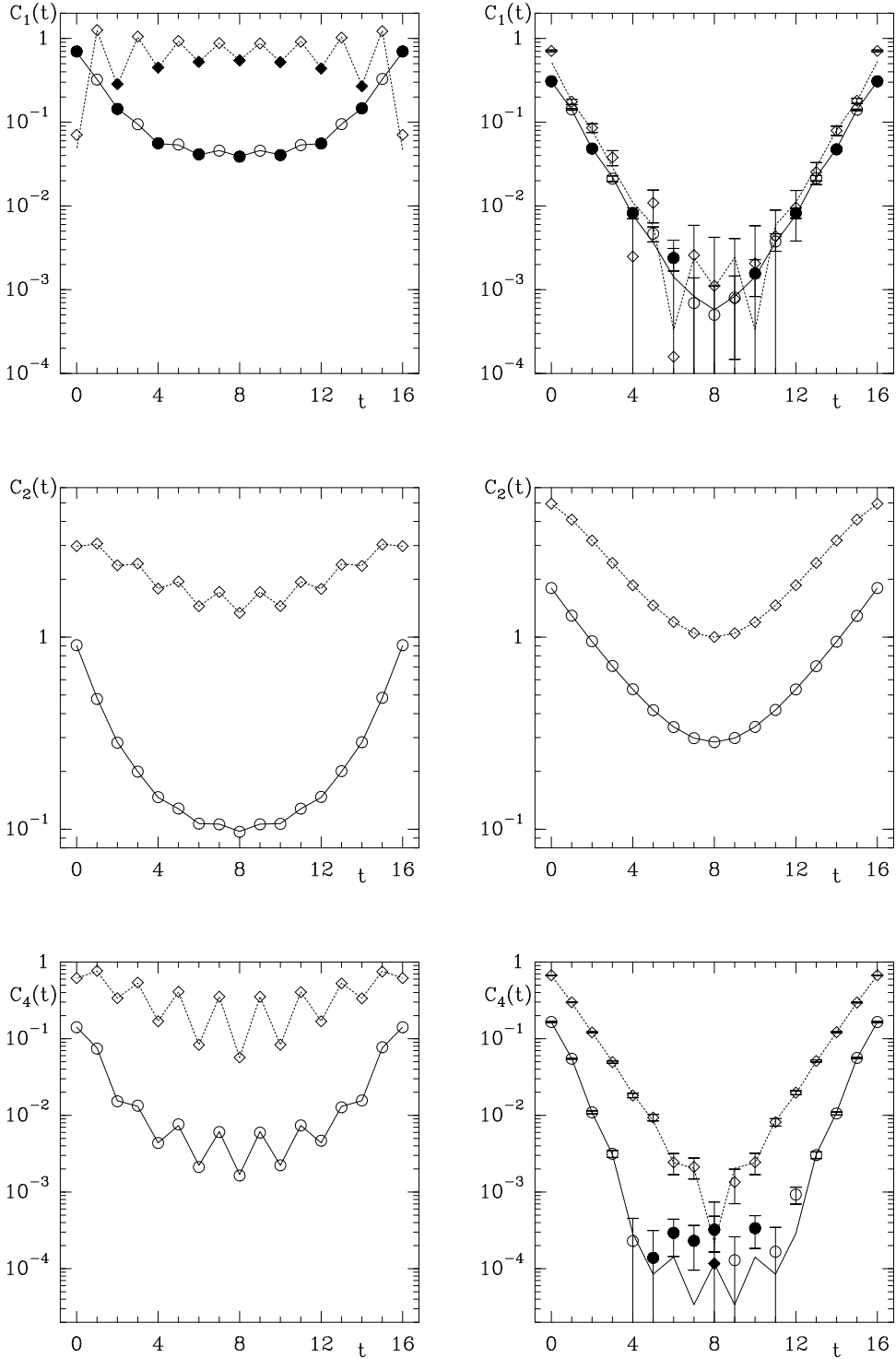
Meson propagators $C_i(t)$ from \MC\ calculations on an $8^3 \times 16$ lattice
at $m_0 = 0.02$ and $g_0 = 0.21$~(l.h.s.), $g_0 = 0.32$ (r.h.s.). Shown are
the local-local propagators $C_i^{(0)}(t)$ (circles) and the wall-local
propagators $C_i^{({\rm w})}(t)$ (diamonds). Open (solid) symbols represent
positive (negative) values. The lines connect values from  simultaneous fits
using \Eq{eq-fit}.}

%------------------------------------------------------------------------
\begin{table}[t]
\begin{center}
\begin{tabular}{|c|ll|lll|}
\hline
\multicolumn{6}{|c|}{$\Epi$} \\
\hline
\hline
& \multicolumn{2}{c|}{$12^4$}&\multicolumn{3}{c|}{$8^3\times 16$}\\
\cline{2-6}
$g_0$&$m_0=0.01$ & $m_0=0.02$ & $m_0=0.02$ & $m_0=0.04$ & $m_0=0.09$\\
\hline
\hline
0.21 & 0.095(4) &          & 0.191(1) & 0.326(1) & 0.582(1) \\
0.23 & 0.129(4) &          & 0.238(1) & 0.377(1) &          \\
0.25 & 0.179(4) & 0.298(2) & 0.280(1) & 0.420(1) &          \\
0.26 &          &          & 0.297(1) & 0.434(1) &          \\
0.27 & 0.238(5) & 0.344(3) & 0.310(1) & 0.442(1) & 0.653(1) \\
0.28 & 0.250(3) &          & 0.317(1) & 0.445(1) &          \\
0.29 &          &          & 0.319(1) & 0.444(1) & 0.644(1) \\
0.30 & 0.244(2) & 0.334(2) & 0.317(1) & 0.442(1) & 0.638(1) \\
0.32 & 0.232(1) &          & 0.311(1) & 0.431(1) & 0.624(1) \\
\hline
\end{tabular}
\end{center}
\begin{center}
\begin{tabular}{|c|ll|lll|}
\hline
\multicolumn{6}{|c|}{$E_{\pi}^{(1)}$} \\
\hline
\hline
& \multicolumn{2}{c|}{$12^4$}&\multicolumn{3}{c|}{$8^3\times 16$}\\
\cline{2-6}
$g_0$&$m_0=0.01$ & $m_0=0.02$ & $m_0=0.02$ & $m_0=0.04$ & $m_0=0.09$\\
\hline
\hline
0.21 & 0.772(5) &          & 0.880(7) & 0.923(6) & 1.024(5) \\
0.23 & 0.678(7) &          & 0.828(10) & 0.885(8) &          \\
0.25 & 0.594(11) & 0.646(8) & 0.807(10) & 0.889(16) &          \\
0.26 &          &          & 0.816(14) & 0.885(11) &          \\
0.27 & 0.570(24) & 0.669(13) & 0.831(21) & 0.947(20) & 1.124(15) \\
0.28 & 0.635(20) &          & 0.866(21) & 0.974(15) &          \\
0.29 &          &          & 0.902(27) & 1.014(11) & 1.232(15) \\
0.30 & 0.800(27) & 0.885(32) & 0.974(33) & 1.064(15) & 1.273(21) \\
0.32 & 0.953(40) &          & 1.096(35) & 1.201(33) & 1.375(22) \\
\hline
\end{tabular}
\end{center}
\CAPTION{tab-pi}{\MC\ results for the energies of the ground state of the \PI\
particle $\Epi$ and the first excited state $E_{\pi}^{(1)}$.}
\end{table}

Let us now turn to the discussion of our fit formulae. By means of a
complete set of eigenstates of the transfer matrix one can write
\begin{eqnarray}
 C(t) & = & \langle {\cal O}(t) {\cal O}'(0) \rangle \nonumber \\
      & = & \sum_{i,j}
 \langle i | \hat{\cal O} | j \rangle
 \langle j | \hat{\cal O}' | i \rangle
 e^{-E_j t - E_i(L_t - t)} \sigma_j^t \sigma_i^{L_t - t}, \label{eq-C}
\end{eqnarray}
with $\sigma_i = \pm 1$. For a more detailed discussion of the
$\hat{\cal O}$ and the $\sigma_i$ see~\cite{Alt92}.
If $L_t$ was infinite one of the intermediate states in the sum would
always be the vacuum,
but on a finite lattice we have a genuine double sum. Terms will occur
in the sum where $E_i = E_j$ which contribute either a constant or an
 ``oscillating constant" depending on whether $\sigma_i$ and $\sigma_j$
 have the same or opposite sign. (The reader may at first be surprised
 to hear that there are states with identical energy but opposite $\sigma$.
 In the fermionic sector there are several discrete symmetries
 which change $\sigma$ but leave $E$ unchanged. For example
 the action (\ref{Eq-S}) is invariant under the transformation
\begin{equation}
 \chi(x) \rightarrow  \zeta_\mu(x) \chi(x+\hat{\mu}),\;
\bar{\chi}(x) \rightarrow  \zeta_\mu(x) \bar{\chi}(x+\hat{\mu}),
\end{equation}
where
\begin{equation}
 \zeta_{\mu}(x) = (-1)^{x_{\mu+1} + \cdots + x_4},\; \zeta_4(x) = 1.
\label{zeta-defn}
\end{equation}
This transformation changes $\sigma$ if $\mu$ is spacelike.
 Such pairs  occur in the fermionic sector because of the well-known
 phenomenon of doubling. The propagator for staggered fermions has
 sixteen states
 instead of the four states of the continuum propagator. Eight of these
 states come from poles with $p_4$ near 0, and so correspond to ordinary
  exponentials in position space, the other eight have $p_4$ near to $\pi$
  and therefore give rise to oscillating exponentials. The above
  symmetries ensure that the decay rate of the
  normal exponentials and the oscillating exponentials remains the same,
  even in the presence of interactions.
 Therefore we know that
 every single fermion state has a partner state with the same energy
 but opposite $\sigma$, so there will indeed be contributions of the
 ``oscillating constant" type in the meson propagator.
 We have already come across a pair of states with the same energy but
 opposite $\sigma$ in the fermion
propagator \Eq{eq-G-fit}, \Fig{fig-prop-mr}. The reader can easily check
 that at small $t$ the propagator $G(t)$ is dominated by a normal
 exponential and at large $t$ by an oscillating, but that in both cases
 the energy is the same.)
Consider two processes contributing to the correlation functions:
\begin{itemize}
\item[(a)] \raisebox{0.618ex}[0.5em][0.5em]{\unitlength 1em
\begin{picture}(2.5,0)(-0.25,0)
\put(0,0){\circle*{0.25}}
\put(2,0){\circle*{0.25}}
\bezier{50}(0,0)(1,1)(2,0)
\bezier{50}(0,0)(1,-1)(2,0)
\put(1.2,+0.5){\vector(+1,0){0}}
\put(0.8,-0.5){\vector(-1,0){0}}
\end{picture}} :
The intermediate state is a fermion-antifermion pair,
\item[(b)] \raisebox{0.618ex}[0.5em][0em]{\unitlength 1em
\begin{picture}(4,0)
\put(1,0){\circle*{0.25}}
\put(3,0){\circle*{0.25}}
\bezier{50}(1,0)(2,1)(3,0)
\put(2.2,+0.5){\vector(+1,0){0}}
\put(0,0){\line(1,0){1}}
\put(3,0){\line(1,0){1}}
\put(0.7,0){\vector(1,0){0}}
\put(3.7,0){\vector(1,0){0}}
\end{picture}} :
A single fermion propagates around the lattice.
\end{itemize}
It is process (b) which gives rise to the constants. This can already be
checked for the free case on a finite lattice. The check is easiest for
the wall propagator $C_i^{({\rm w})}$ which in the free theory is simply the
square of the fermion

%------------------------------------------------------------------------
\begin{table}[p]%
\begin{center}%
\begin{tabular}{|c|ll|lll|}%
\hline
\multicolumn{6}{|c|}{$\fpi$} \\
\hline
\hline
& \multicolumn{2}{c|}{$12^4$}&\multicolumn{3}{c|}{$8^3\times 16$}\\
\cline{2-6}
$g_0$&$m_0=0.01$ & $m_0=0.02$ & $m_0=0.02$ & $m_0=0.04$ & $m_0=0.09$\\
\hline
\hline
0.21 & 0.073(5) &          & 0.115(1) & 0.107(1) & 0.109(0) \\
0.23 & 0.070(3) &          & 0.111(1) & 0.110(1) &          \\
0.25 & 0.071(3) & 0.076(1) & 0.112(1) & 0.116(1) &          \\
0.26 &          &          & 0.116(1) & 0.124(1) &          \\
0.27 & 0.084(3) & 0.095(1) & 0.120(1) & 0.128(1) & 0.147(1) \\
0.28 & 0.096(2) &          & 0.126(1) & 0.135(1) &          \\
0.29 &          &          & 0.131(1) & 0.141(0) & 0.160(0) \\
0.30 & 0.119(1) & 0.129(2) & 0.139(1) & 0.147(1) & 0.164(1) \\
0.32 & 0.141(1) &          & 0.151(1) & 0.160(1) & 0.173(0) \\
\hline
\end{tabular}%
\end{center}%
\CAPTION{tab-fpi}{\MC\ results for the pion decay constant $\fpi$.}%
\end{table}%
\samepage%
%------------------------------------------------------------------------
\begin{table}[p]%
\begin{center}%
\begin{tabular}{|c|ll|lll|}%
\hline
\multicolumn{6}{|c|}{$E_{\pi'}^{(0)}$} \\
\hline
\hline
& \multicolumn{2}{c|}{$12^4$}&\multicolumn{3}{c|}{$8^3\times 16$}\\
\cline{2-6}
$g_0$&$m_0=0.01$ & $m_0=0.02$ & $m_0=0.02$ & $m_0=0.04$ & $m_0=0.09$\\
\hline
\hline
0.21 & 0.100(7) &          & 0.272(4) & 0.430(3) & 0.705(2) \\
0.23 & 0.145(7) &          & 0.361(9) & 0.522(6) &          \\
0.25 & 0.227(9) & 0.380(5) & 0.483(12) & 0.645(8) &          \\
0.26 &          &          & 0.541(17) & 0.687(9) &          \\
0.27 & 0.377(22) & 0.529(15) & 0.617(19) & 0.789(13) & 1.032(8) \\
0.28 & 0.501(25) &          & 0.675(26) & 0.830(15) &          \\
0.29 &          &          & 0.747(33) & 0.907(11) & 1.130(12) \\
0.30 & 0.683(53) & 0.783(25) & 0.796(44) & 0.981(28) & 1.223(16) \\
0.32 & 0.892(110) &          & 0.934(72) & 1.131(45) & 1.303(24) \\
\hline
\end{tabular}%
\end{center}%
\begin{center}%
\begin{tabular}{|c|ll|lll|}%
\hline
\multicolumn{6}{|c|}{$E_{\sigma}^{(0)}$} \\
\hline
\hline
& \multicolumn{2}{c|}{$12^4$}&\multicolumn{3}{c|}{$8^3\times 16$}\\
\cline{2-6}
$g_0$&$m_0=0.01$ & $m_0=0.02$ & $m_0=0.02$ & $m_0=0.04$ & $m_0=0.09$\\
\hline
\hline
0.21 & 0.751(8) &          & 0.877(6) & 0.939(4) & 1.105(3) \\
0.23 & 0.665(9) &          & 0.817(6) & 0.922(4) &          \\
0.25 & 0.595(11) & 0.680(6) & 0.811(5) & 0.936(9) &          \\
0.26 &          &          & 0.816(7) & 0.940(5) &          \\
0.27 & 0.598(17) & 0.705(11) & 0.824(11) & 0.974(12) & 1.206(10) \\
0.28 & 0.656(15) &          & 0.855(11) & 0.996(8) &          \\
0.29 &          &          & 0.854(14) & 1.026(6) & 1.271(9) \\
0.30 & 0.771(25) & 0.873(15) & 0.880(17) & 1.049(23) & 1.300(19) \\
0.32 & 0.898(47) &          & 0.926(29) & 1.152(21) & 1.364(17) \\
\hline
\end{tabular}%
\end{center}%
\CAPTION{tab-sigma}{\MC\ results for the energies of ground state of the
$\pi'$ particle $E_{\pi'}^{(0)}$ and
the ground state energy of the \SIGMA\
particle $E_{\sigma}^{(0)}$.}%
\end{table}%
\noindent propagator $G(t)$ (see \Eq{eq-G-fit})
\begin{equation}
 C_{\rm free}^{({\rm w})}(t) = G_{\rm free}^2(t) \propto
 e^{-2\mR t} + e^{-2\mR (L_t - t)} - 2 (-1)^t e^{-\mR L_t}
\end{equation}
which already displays an oscillating constant contribution.

 From \Eq{eq-C} $C(t)$ thus has the form
\begin{eqnarray}
 C(t) & = & \sum_n A_n^{+}
            \left( e^{-E_n^{+} t} + e^{-E_n^{+} (L_t - t)} \right) \nonumber \\
      & + & \sum_n A_n^{-}
            \left( e^{-E_n^{-} t} + e^{-E_n^{-} (L_t - t)} \right)
            (-1)^t \nonumber \\
      & + & B^{+} \nonumber \\
      & + & B^{-} (-1)^t \,. \label{eq-fit}
\end{eqnarray}
The $E^{+}$ terms describe $s$-wave states and the $E^{-}$ terms describe the
parity partners of these states which are $p$-waves. The quantum numbers
are given in \Tab{tab-names}. In the following the energies are called
$E_{\rm name}^{(i)}$ where ``name'' is the particle name from \Tab{tab-names}
and $i = 0$ or~1 for the energies of the ground state or first excited state.

In fitting we had to restrict ourselves to the lightest states.
We used two $s$-wave states for the \PI\ and  one $s$- and
one $p$-wave state for the $\pi'/\sigma$. For all vector channels
  ($\rho'/a$, $\rho/b$, $\omega/b'$) we used
 two $s$-wave states and one
$p$-wave state. The choices of the terms in the
fit function were necessary to obtain good fits. In the case of the
\PI\ correlation function it is further motivated by the fact that a
$0^{+-}$ state cannot be realised as a fermion-antifermion state. In
the other cases we were guided by the results from the \SDeqs.
It was found that $B^{+}$ was always negligible and that
$B^{-}$ was necessary to describe propagators in the symmetric phase.
In order to
extract a maximum of information from our correlation functions we found
it convenient to make a simultaneous fit to both the local-local and the
wall-local correlation functions.

%------------------------------------------------------------------------
\begin{table}[p]
\begin{center}
\begin{tabular}{|c|ll|lll|}
\hline
\multicolumn{6}{|c|}{$E_{\rho}^{(0)}$} \\
\hline
\hline
& \multicolumn{2}{c|}{$12^4$}&\multicolumn{3}{c|}{$8^3\times 16$}\\
\cline{2-6}
$g_0$&$m_0=0.01$ & $m_0=0.02$ & $m_0=0.02$ & $m_0=0.04$ & $m_0=0.09$\\
\hline
\hline
0.21 & 0.101(2) &          & 0.253(1) & 0.407(1) & 0.678(1) \\
0.23 & 0.144(2) &          & 0.344(2) & 0.499(2) &          \\
0.25 & 0.227(3) & 0.371(2) & 0.460(3) & 0.620(2) &          \\
0.26 &          &          & 0.526(4) & 0.677(3) &          \\
0.27 & 0.374(5) & 0.521(4) & 0.595(5) & 0.749(4) & 0.992(4) \\
0.28 & 0.485(5) &          & 0.662(7) & 0.817(6) &          \\
0.29 &          &          & 0.741(9) & 0.883(4) & 1.111(5) \\
0.30 & 0.676(11) & 0.776(7) & 0.829(13) & 0.935(11) & 1.187(11) \\
0.32 & 0.890(29) &          & 0.969(43) & 1.076(17) & 1.290(19) \\
\hline
\end{tabular}
\end{center}
\begin{center}
\begin{tabular}{|c|ll|lll|}
\hline
\multicolumn{6}{|c|}{$E_{\rho}^{(1)}$} \\
\hline
\hline
& \multicolumn{2}{c|}{$12^4$}&\multicolumn{3}{c|}{$8^3\times 16$}\\
\cline{2-6}
$g_0$&$m_0=0.01$ & $m_0=0.02$ & $m_0=0.02$ & $m_0=0.04$ & $m_0=0.09$\\
\hline
\hline
0.21 & 1.36(1) &          & 1.40(2) & 1.46(2) & 1.54(1) \\
0.23 & 1.34(1) &          & 1.42(2) & 1.48(2) &          \\
0.25 & 1.30(2) & 1.35(1) & 1.47(2) & 1.47(4) &          \\
0.26 &          &          & 1.38(3) & 1.43(2) &          \\
0.27 & 1.34(4) & 1.34(3) & 1.39(4) & 1.61(7) & 1.56(3) \\
0.28 & 1.52(7) &          & 1.52(7) & 1.44(3) &          \\
0.29 &          &          & 1.42(6) & 1.43(2) & 1.57(3) \\
0.30 & 1.57(11) & 1.46(5) & 1.52(9) & 1.51(10) & 1.58(6) \\
0.32 & 1.32(10) &          & 1.21(9) & 1.60(10) & 1.53(5) \\
\hline
\end{tabular}
\end{center}
\begin{center}
\begin{tabular}{|c|ll|lll|}
\hline
\multicolumn{6}{|c|}{$E_{b}^{(0)}$} \\
\hline
\hline
& \multicolumn{2}{c|}{$12^4$}&\multicolumn{3}{c|}{$8^3\times 16$}\\
\cline{2-6}
$g_0$&$m_0=0.01$ & $m_0=0.02$ & $m_0=0.02$ & $m_0=0.04$ & $m_0=0.09$\\
\hline
\hline
0.21 & 1.41(3) &          & 1.49(5) & 1.65(5) & 1.73(3) \\
0.23 & 1.42(4) &          & 1.48(6) & 1.69(6) &          \\
0.25 & 1.35(5) & 1.46(4) & 1.63(7) & 1.64(11) &          \\
0.26 &          &          & 1.52(8) & 1.63(6) &          \\
0.27 & 1.53(15) & 1.40(10) & 1.51(13) & 1.97(20) & 1.77(11) \\
0.28 & 1.98(24) &          & 1.97(22) & 1.66(9) &          \\
0.29 &          &          & 1.55(15) & 1.62(6) & 1.75(8) \\
0.30 & 1.90(31) & 1.60(13) & 1.72(21) & 1.99(34) & 1.77(14) \\
0.32 & 1.23(19) &          & 1.45(21) & 2.11(29) & 1.82(11) \\
\hline
\end{tabular}
\end{center}
\CAPTION{tab-rho}{\MC\ results for the energies of the ground state
of the \RHO\ particle $E_{\rho}^{(0)}$, its first exited state $E_{\rho}^{(1)}$
and the ground state of the $b$
particle $E_{b}^{(0)}$.}
\end{table}
%------------------------------------------------------------------------
\begin{table}[p]
\begin{center}
\begin{tabular}{|c|ll|lll|}
\hline
\multicolumn{6}{|c|}{$E_{\rho'}^{(0)}$} \\
\hline
\hline
& \multicolumn{2}{c|}{$12^4$}&\multicolumn{3}{c|}{$8^3\times 16$}\\
\cline{2-6}
$g_0$&$m_0=0.01$ & $m_0=0.02$ & $m_0=0.02$ & $m_0=0.04$ & $m_0=0.09$\\
\hline
\hline
0.21 & 0.100(2) &          & 0.263(2) & 0.418(2) & 0.692(2) \\
0.23 & 0.140(2) &          & 0.352(3) & 0.509(3) &          \\
0.25 & 0.224(3) & 0.374(2) & 0.470(4) & 0.633(3) &          \\
0.26 &          &          & 0.534(5) & 0.688(4) &          \\
0.27 & 0.380(6) & 0.519(5) & 0.608(7) & 0.770(5) & 1.016(5) \\
0.28 & 0.490(7) &          & 0.680(10) & 0.830(7) &          \\
0.29 &          &          & 0.767(12) & 0.904(5) & 1.141(7) \\
0.30 & 0.685(14) & 0.784(10) & 0.837(16) & 0.943(12) & 1.214(13) \\
0.32 & 0.840(25) &          & 0.930(31) & 1.099(17) & 1.282(11) \\
\hline
\end{tabular}
\end{center}
\begin{center}
\begin{tabular}{|c|ll|lll|}
\hline
\multicolumn{6}{|c|}{$E_{\rho'}^{(1)}$} \\
\hline
\hline
& \multicolumn{2}{c|}{$12^4$}&\multicolumn{3}{c|}{$8^3\times 16$}\\
\cline{2-6}
$g_0$&$m_0=0.01$ & $m_0=0.02$ & $m_0=0.02$ & $m_0=0.04$ & $m_0=0.09$\\
\hline
\hline
0.21 & 1.40(3) &          & 1.46(5) & 1.61(4) & 1.68(2) \\
0.23 & 1.41(4) &          & 1.48(6) & 1.61(5) &          \\
0.25 & 1.40(5) & 1.44(4) & 1.64(7) & 1.60(9) &          \\
0.26 &          &          & 1.66(10) & 1.63(5) &          \\
0.27 & 1.48(8) & 1.64(10) & 1.51(13) & 1.76(13) & 1.80(8) \\
0.28 & 1.50(12) &          & 1.51(12) & 1.61(7) &          \\
0.29 &          &          & 2.08(33) & 1.93(9) & 1.90(7) \\
0.30 & 1.78(26) & 1.58(11) & 2.04(34) & 1.70(20) & 1.68(11) \\
0.32 & 2.07(57) &          & 1.70(27) & 2.12(31) & 2.23(18) \\
\hline
\end{tabular}
\end{center}
\begin{center}
\begin{tabular}{|c|ll|lll|}
\hline
\multicolumn{6}{|c|}{$E_{a}^{(0)}$} \\
\hline
\hline
& \multicolumn{2}{c|}{$12^4$}&\multicolumn{3}{c|}{$8^3\times 16$}\\
\cline{2-6}
$g_0$&$m_0=0.01$ & $m_0=0.02$ & $m_0=0.02$ & $m_0=0.04$ & $m_0=0.09$\\
\hline
\hline
0.21 & 1.36(1) &          & 1.43(2) & 1.48(1) & 1.57(1) \\
0.23 & 1.34(1) &          & 1.42(2) & 1.50(2) &          \\
0.25 & 1.32(2) & 1.36(1) & 1.53(2) & 1.50(3) &          \\
0.26 &          &          & 1.49(3) & 1.49(2) &          \\
0.27 & 1.38(3) & 1.46(4) & 1.46(4) & 1.61(5) & 1.71(3) \\
0.28 & 1.40(4) &          & 1.46(4) & 1.54(3) &          \\
0.29 &          &          & 1.75(11) & 1.65(3) & 1.75(3) \\
0.30 & 1.51(8) & 1.48(4) & 1.66(10) & 1.61(8) & 1.70(4) \\
0.32 & 1.79(22) &          & 1.49(8) & 1.79(10) & 1.88(6) \\
\hline
\end{tabular}
\end{center}
\CAPTION{tab-a}{\MC\ results for the energies of
the ground state of the $\rho'$
particle $E_{\rho'}^{(0)}$, its first exited state $E_{\rho'}^{(1)}$
and the ground state energy of the $a$ particle
$E_{a}^{(0)}$.}
\end{table}
%------------------------------------------------------------------------
\begin{table}[p]
\begin{center}
\begin{tabular}{|c|ll|lll|}
\hline
\multicolumn{6}{|c|}{$E_{\omega}^{(0)}$} \\
\hline
\hline
& \multicolumn{2}{c|}{$12^4$}&\multicolumn{3}{c|}{$8^3\times 16$}\\
\cline{2-6}
$g_0$&$m_0=0.01$ & $m_0=0.02$ & $m_0=0.02$ & $m_0=0.04$ & $m_0=0.09$\\
\hline
\hline
0.21 & 0.22(4) &          & 0.27(1) & 0.43(1) & 0.71(1) \\
0.23 & 0.30(4) &          & 0.37(2) & 0.53(1) & 0.80(2) \\
0.25 & 0.46(5) & 0.47(4) & 0.50(4) & 0.64(3) & 0.94(4) \\
0.26 &          &          & 0.62(4) & 0.71(4) & 1.00(6) \\
0.27 & 0.56(13) & 0.40(15) & 0.82(8) & 0.78(5) & 1.11(5) \\
0.28 & 0.54(12) &          & 0.58(9) & 0.80(8) & 1.01(14) \\
0.29 &          &          & 0.80(12) & 1.04(7) & 1.32(8) \\
0.30 & 0.99(10) & 1.03(4) & 1.29(9) & 0.91(21) & 1.29(15) \\
0.32 & 1.22(11) &          & 0.97(20) & 0.97(15) & 1.18(25) \\
\hline
\end{tabular}
\end{center}
\begin{center}
\begin{tabular}{|c|ll|lll|}
\hline
\multicolumn{6}{|c|}{$E_{\omega}^{(1)}$} \\
\hline
\hline
& \multicolumn{2}{c|}{$12^4$}&\multicolumn{3}{c|}{$8^3\times 16$}\\
\cline{2-6}
$g_0$&$m_0=0.01$ & $m_0=0.02$ & $m_0=0.02$ & $m_0=0.04$ & $m_0=0.09$\\
\hline
\hline
0.21 & 1.68(3) &          & 1.75(4) & 1.76(3) & 1.78(2) \\
0.23 & 1.74(6) &          & 1.68(7) & 1.75(5) & 1.81(3) \\
0.25 & 1.97(18) & 1.72(6) & 1.79(18) & 1.78(6) & 1.90(6) \\
0.26 &          &          & 1.86(14) & 1.86(11) & 1.90(7) \\
0.27 & 1.82(34) & 1.57(10) & 2.23(51) & 1.85(11) & 2.05(11) \\
0.28 & 1.68(21) &          & 1.64(10) & 1.80(9) & 1.88(9) \\
0.29 &          &          & 1.80(22) & 2.22(27) & 2.28(27) \\
0.30 &          &          &          & 1.74(12) & 2.12(25) \\
0.32 &          &          &          & 2.13(31) & 1.98(15) \\
\hline
\end{tabular}
\end{center}
\begin{center}
\begin{tabular}{|c|ll|lll|}
\hline
\multicolumn{6}{|c|}{$E_{b'}^{(0)}$} \\
\hline
\hline
& \multicolumn{2}{c|}{$12^4$}&\multicolumn{3}{c|}{$8^3\times 16$}\\
\cline{2-6}
$g_0$&$m_0=0.01$ & $m_0=0.02$ & $m_0=0.02$ & $m_0=0.04$ & $m_0=0.09$\\
\hline
\hline
0.21 & 1.55(4) &          & 1.67(7) & 1.75(5) & 1.71(3) \\
0.23 & 1.67(7) &          & 1.42(18) & 1.64(13) & 1.67(8) \\
0.25 & 1.90(18) & 1.64(7) & 1.83(48) & 1.70(8) & 1.79(11) \\
0.26 &          &          & 1.57(23) & 1.98(30) & 1.91(16) \\
0.27 & 1.69(34) & 1.56(14) & 2.15(46) & 1.82(13) & 1.93(10) \\
0.28 & 1.49(21) &          & 1.27(26) & 1.68(23) & 1.70(16) \\
0.29 &          &          & 1.69(24) & 2.15(22) & 2.15(19) \\
0.30 &          &          &          & 1.61(14) & 2.04(17) \\
0.32 &          &          &           & 2.33(46) & 2.02(19) \\
\hline
\end{tabular}
\end{center}
\CAPTION{tab-omega}{\MC\ results for the energies of
ground state of the $\omega$ particle $E_{\omega}^{(0)}$,
its first exited state $E_{\omega}^{(1)}$ and
the ground state of the $b'$ particle $E_{b'}^{(0)}$.}
\end{table}

\hfig{p}{fig-anni-pi-mc}{%
Influence of the annihilation part on the \PI\ propagator in the \MC\
calculation.
Parameters and symbols are the same as in \Fig{fig-prop-mc} except that the
diamonds now represent the complete propagator, \Eq{eq-picture}.}

\hfig{p}{fig-anni-sigma-mc}{%
Influence of the annihilation part on the $\pi' / \sigma$ propagator in
the \MC\ calculation.
Parameters and symbols are the same as in \Fig{fig-prop-mc} except that the
diamonds now represent the complete propagator, \Eq{eq-picture}. See
text for comments on \bbb.}

{}From the \PI\ ground state energy $\Epi$ and amplitude $A_\pi^{(0)}$
we calculated the pion decay constant $\fpi$ using~\cite{Kil87}:
\begin{equation}
 \fpi = \frac{m_0}{4} \sqrt{\left| A_\pi^{(0)} \right|} \,
        \frac{\sqrt{\sinh \Epi}}{\sinh^2(\Epi / 2)} \,.
\end{equation}
The fit results are given in tables~\ref{tab-pi}--\ref{tab-omega}. As we have
seen for the chiral condensate the finite size effects are quite large. In
\Fig{fig-prop-mc} we show propagators and fits of the $\pi'/\sigma$, $\pi$ and
$\rho/b$ (top to bottom) for one $g_0$-value in the symmetric and one
$g_0$-value in the broken phase.

For certain values of $g_0$ and $m_0$ we have also measured the
annihilation part of the \PI\ and \SIGMA\ correlation functions, where
16 sets of gaussian random numbers were used in the stochastic estimator. As
\Fig{fig-anni-pi-mc} shows the influence in the \PI\ correlation
function is small. Since the \SIGMA\ has the quantum numbers of the
vacuum we have to subtract a constant from the \SIGMA\ correlation
function before comparing it with the result without the annihilation
part. As the constant is rather large (it is of $O(L_s^3 \CHI^2)$)
 in the middle of the
correlation function in the broken phase we only see noise.
A typical result is given in \Fig{fig-anni-sigma-mc}. Within the
errors the energies are unchanged and thus we can use the numbers in
\Tab{tab-sigma} with some confidence.
The discussion of our results will be postponed until section
\ref{sect-spectroscopy}.

\subsection{\SD\ Equations}

\PSfigure{b}{(14.1,8.0)}{(0,1.6)}{fig-sd-c}{fig-sd-c}{%
Pictorial representation of the \SDeq\ for the meson propagators.}

 \noindent  The fact that the \SDeqs\ have given
  such good results for the fermion propagator and chiral
  condensate encourages us to try a similar approach to meson
  propagators.  In \Fig{fig-sd-c} we show the  \SDeqs\
  for the fermion-antifermion composite
  propagator.

  To get a manageable set of equations we have
  replaced the full kernel with the bare 4-fermi interaction.
  This gives us equations which are at the same level of accuracy
as the order $g_0$ \SD\ equations for the fermion propagator which, as
  we have already seen, yield a good approximation to the \MC\ results.

For the fermion line connected part of the local correlation functions we
derive the following results in appendix \ref{app-SDeqs}. Defining
\begin{equation}
 C_i^{(0)}(t) =
    \frac{1}{L_t} \sum_\omega e^{i\omega t} \tilde{C}_i^{(0)}(\omega)\,,
    \label{eq-fourier}
\end{equation}
we find that the Fourier transforms $\tilde{C}_i^{(0)}(\omega)$ are given by

\begin{eqnarray}
\tilde{C}_1^{(0)}(\omega) & = &\frac{a_1(\omega)}
    {1 + a_1(\omega) 2 g_0 (+3 - \cos \omega) } \,, \nonumber \\  %scalar
\tilde{C}_2^{(0)}(\omega) & = &\frac{a_2(\omega)}
    {1 + a_2(\omega) 2 g_0 (-3 - \cos \omega) } \,, \nonumber \\  %pion
\tilde{C}_3^{(0)}(\omega) & = &\frac{a_3(\omega)}
    {1 + a_3(\omega) 2 g_0 (+1 - \cos \omega) } \,, \nonumber \\  %vec2
\tilde{C}_4^{(0)}(\omega) & = &\frac{a_4(\omega)}
    {1 + a_4(\omega) 2 g_0 (-1 - \cos \omega) } \,, \label{eq-C-tilde} %vec1
\end{eqnarray}

\noindent
where
\begin{equation}
   a_i(\omega) =
     \frac{1}{V} \sum_k
     \frac{ B_i^2 + F_t^2 \sin k_4 \sin(k_4 -\omega) }
     { \left[ A^2 + F_t^2 \sin^2 k_4 \right]
       \left[ A^2 + F_t^2 \sin^2 (k_4-\omega) \right] } \,,
    \label{aiom}
\end{equation}

\noindent
with
\begin{eqnarray}
 A^2   & \equiv & N^2 + F_s^2 s^2 \,, \nonumber \\
 B_1^2 & \equiv & N^2 - F_s^2 s^2 \,, \nonumber \\
 B_2^2 & \equiv & N^2 + F_s^2 s^2 \,, \nonumber \\
 B_3^2 & \equiv & N^2 -\frac{1}{3} F_s^2 s^2 \,, \nonumber \\
 B_4^2 & \equiv & N^2 +\frac{1}{3} F_s^2 s^2 \,, \label{a-and-b}
\end{eqnarray}

\noindent
and
\begin{equation}
 s^2 \equiv \sum_{i=1}^3 \sin^2 k_i \,.
\end{equation}

\noindent
The $N$, $F_s = F_1 = F_2 = F_3$
and $F_t = F_4$ have been defined in (\ref{OGfermi}).
The formulae for the fermion line disconnected part are also given in appendix
\ref{app-SDeqs}.

     The principal interest in the fermion-antifermion channels lies in
   the energy spectrum, i.~e., the eigenvalues of the transfer matrix.
   These energies can best be
   found on a long lattice which is why we used $8^3 \times 16$
   lattices for most of our simulations. In the
   \SD\ approach we can easily extend $L_t$ to infinity,
   which allows a clean extraction of the full spectrum.

       When $L_t = \infty$ the $k_4$ sum becomes an integral,
   which can be evaluated by contour integration.  The
   general ``bubble" $a_i$ in (\ref{aiom}) has the form
\begin{eqnarray}
 a_i(\omega) & = & \frac{1}{L_s^3} \sum_{\vec{k}} {\cal I}(A, B, \omega) \,,
             \label{eq-ai} \\
 {\cal I}(A,B,\omega) & = &
     \int_0^{2\pi} \frac{d k_4}{2 \pi}
     \frac{ B^2 + F_t^2 \sin k_4 \sin(k_4 -\omega) }
     { \left[ A^2 + F_t^2 \sin^2 k_4 \right]
       \left[ A^2 + F_t^2 \sin^2 (k_4-\omega) \right] } \,,\,\,
\end{eqnarray}
\noindent
where $A$ and $B$ are functions of $\vec{k}$
only. Evaluating this integral leads to the result
\begin{equation}
    {\cal I}(A,B,\omega) =
      \frac{1}{A \sqrt{A^2+F_t^2} } \left(
     \frac{  A^2 + B^2}
     {  2 A^2 + F_t^2 -F_t^2 \cos \omega }
      - \frac{  A^2 - B^2}
      { 2 A^2 + F_t^2 +F_t^2 \cos \omega } \right).
     \label{calI}
\end{equation}
     {}From the form of ${\cal I}$ it is easy to see that the
  Fourier transform of the composite propagator is the
  ratio of two real polynomials in $\cos \omega$.
   Therefore the only singularities that can occur
  when $L_s$ is finite are poles.
  A pole in the momentum-space propagator at
  $\omega = i E$ corresponds to a  term \mbox{$\propto \exp(-E t)$}
   in the real-space propagator, and a pole at
  $\omega = \pi + i E$ to a term \mbox{$\propto (-1)^t \exp(-E t)$},
  so we can find the complete energy spectrum by locating
   all the poles in the momentum-space propagator.
  The order of these
  polynomials is proportional to the number of distinct terms
  in the sum over spatial momenta, so the number of energy
  levels should grow as $L_s^3$.

\PSfigure{t}{(14.1,10.0)}{(0,1)}{fig-cuts}{fig-cuts}{%
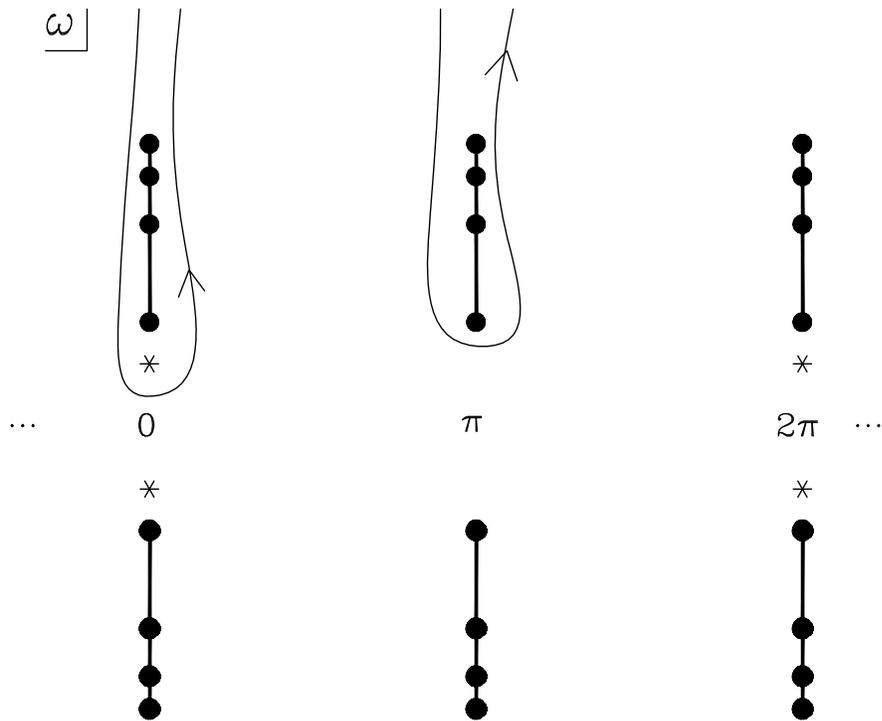
Singularity structure of $\tilde{C}_i^{(0)}(\omega)$ in the $\omega$-plane.
The solid lines are cuts, a $\ast$ is a pole and
 a $\bullet$ is a branch point.}

Cursory inspection of the meson levels in
tables~\ref{tab-pi}--\ref{tab-omega} shows that many of the
energy levels lie close to or above $2\mR$.
     In order to really understand what is happening above the
  fermion-antifermion threshold we need to take the infinite volume
  limit and see what happens as the excited states form a
  true continuum.
We have already taken the $L_t \rightarrow \infty$ limit in \Eq{calI}. We now
let $L_s \rightarrow \infty$. The sum over $\vec{k}$ in \Eq{aiom} then
turns into an integral, the analytic structure of which is worked out in
appendix \ref{app-inf-vol}. As $L_s$ increases the poles above the threshold
become denser and in the infinite volume limit become a cut.

In \Fig{fig-cuts} we sketch the singularity structure of $\tilde{C}$ in the
$\omega$-plane and also give the integration contour needed to evaluate $C(t)$.
  $\tilde{C}(\omega)$ has cuts running
 from $\cos \omega = 1 + 2  \sinh^2 \mR$ to
 $\cos \omega = 7 + 2  \sinh^2 \mR$  and
 from $\cos \omega = -1 - 2  \sinh^2 \mR$ to
 $\cos \omega =- 7 - 2  \sinh^2 \mR$.  These cuts represent
  the continuum of fermion-antifermion states running from
  $E = 2 \mR$ to $E =  \cosh^{-1}(7 + 2 \sinh^2 \mR )$,
  the highest energy available to two fermions.  The singularities
  in the phase-space function $\phi(r)$ (see appendix B)
  at $r=\pm1$ lead to additional
  branch points within the cuts at
  $\cos \omega = \pm(3 + 2 \sinh^2 \mR)$ and
  $\cos \omega = \pm(5 + 2 \sinh^2 \mR)$.

  The discontinuity across the cuts of $\tilde{C}$ gives the spectral
  function $\rho$~\cite{ELOP}. As well as these cuts $\tilde{C}$ can also
  exhibit poles, which correspond to genuine bound states (stable mesons).

\subsection{Finite Size Effects}

We shall now study \FS\ effects for bound state masses using methods similar to
those used in section \ref{sect-fs-chi} for the chiral condensate and the
fermion mass. To do this we will consider \FS\ corrections to the ``bubble''
sums given in \Eq{eq-M-bubbles}. The main characteristic of these bubbles is
that they involve two propagators with denominators $U^{+}$ and $U^{-}$. We
take a slightly  more general case where the physical masses ($m_+$ and $m_-$)
in the two propagators can be different.

To find the change in the meson propagator we must investigate the effect of
replacing the sum over loop momenta by an integral. Since we are interested in
the meson mass we need to consider the case where the meson
is \, on-shell, i.~e., \,
the 4-momentum $p = (0,0,0, i M)$ where \, $M$ \, is the meson
\clearpage
\noindent
 mass. We now use the
same Fourier transform argument as in section \ref{sect-fs-chi} to change a sum
in momentum space to a sum in coordinate space.

\PSfigure{b}{(12,10)}{(0,1)}{fig-lorentz}{fig-lorentz}{%
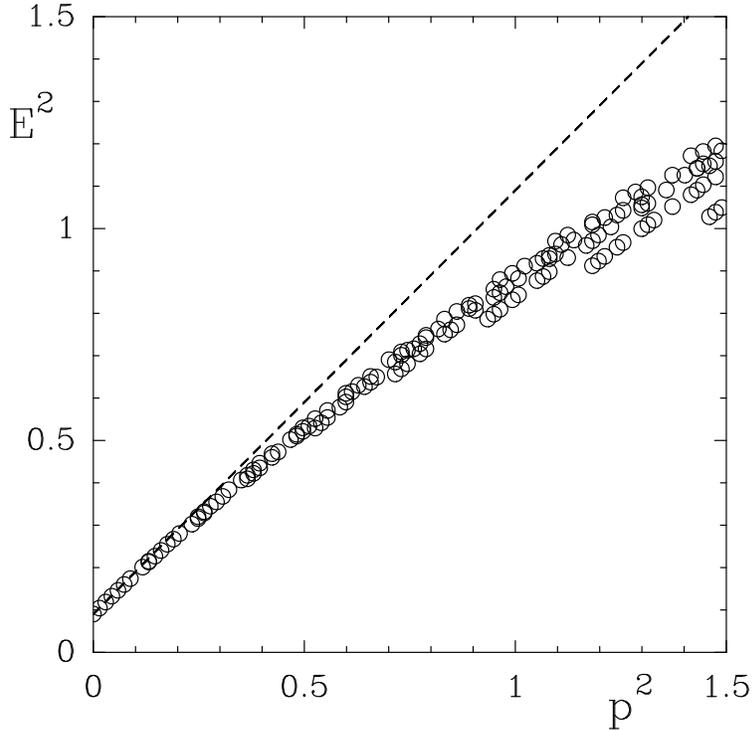
Dispersion relation of the \PI\ bound state obtained from \SD\ calculations.
Results are from a $52^3 \times \infty$ lattice with
$m_0=0.01$ and $g_0=0.2476$, leading to $\mu_R=0.20$ and $m_{\pi}=0.301$.}

A typical loop integral in eq.~(\ref{eq-M-bubbles}) has the form
\begin{equation}
 \int \frac{ d^4 k}{(2\pi)^4}
   \frac{ n(p,k) }{ \left( (p+k)^2 + m_+^2 \right)
                   \left( k^2 + m_-^2 \right) }  \ .
 \end{equation}
(The exact form of the numerator $n(p,k)$ is unimportant for  asymptotic
behaviour.)  As before we need the large $L_s$ limit  of
\begin{equation}
 \int \frac{ d^4 k}{(2\pi)^4} \exp (i k_1 L_s )
   \frac{ n(p,k) }{ \left( (p+k)^2 + m_+^2 \right)
                   \left( k^2 + m_-^2 \right) }  \ .
 \end{equation}
   Integrating exactly over $k_1$ and then with the saddle point
 approximation over the other three components of $k$ gives a result
\begin{equation}
 \propto \frac{1}{L_s} \exp( -\xi L_s ) \equiv \Delta,
\label{meson-shift}
\end{equation}
 where
\begin{equation}
 \xi =  \frac{ \sqrt{ -M^4 + 2 M^2 m_+^2 + 2 M^2 m_-^2
    + 2 m_+^2 m_-^2 - m_+^4 - m_-^4 }}{2 M} \ .
\label{xi-forml}
\end{equation}
  The finite-size shift in the meson mass ought to be proportional
 to $\Delta$.

  We can consider various special cases of eq.~(\ref{xi-forml}).
 In the non-relativistic limit where the binding energy
 $ B \equiv m_+ + m_- - M $ is small compared with $M$
\begin{equation}
 \xi \to \sqrt{ 2 m_{\mbox{\scriptsize reduced}} B} \equiv
\sqrt{ 2 \left(\frac{m_+ m_-}{ m_+ +m_-}\right)
\left( m_+ + m_- - M \right)  },
\end{equation}
recovering a result that can be found by solving the Schr\"odinger equation
in a finite volume.

Other useful limits are the case $m_+ = m_-$ which gives     $\xi = (4 m_+^2
- M^2)^{1/2}/2 $, which should apply to the   pion when it is a true bound
state,    and $M = m_+ = m_-$ which gives    $\xi = \frac{\sqrt{3}}{2} M $,
relevant for finding the pion mass shift due to $\pi \to 2\pi \to
\pi$~\cite{Lus91}.  A further application of eq.~(\ref{meson-shift}) is to the
process $ f \to f + \pi \to f $, which will give a finite size shift to the
fermion mass $\mu_R$ additional to that considered  in section
\ref{sect-fs-chi}. This extra contribution will be of the form
(\ref{meson-shift}) with    $\xi = \frac{m_\pi}{2 \mR} \sqrt{4 \mR^2 -
m_\pi^2}$ and can become the leading term if the pion is light enough.

With the \SDeqs\ it is easy to study the restoration of Lorentz invariance.
As an example
in \Fig{fig-lorentz} we look at the dispersion relation of the
  \PI\ bound state. We have plotted
 $E^2$ against
 $\vec{p}\,^2$ for all possible
  $\vec{p}$ values which occur on a lattice with $L_s=52$.
  When $\vec{p}\,^2$ can be realised in inequivalent ways we have plotted
 all the possible energies. We see that the Einstein relationship
  $E^2 = m^2 + \vec{p}\,^2$ holds well up to $\vec{p}\,^2 \approx 0.2$
  and that spherical symmetry ($E^2$ independent of the momentum
  direction) holds up to  $\vec{p}\,^2 \approx 0.5$.

Although Lorentz invariance is restored we observe at most partial
flavour symmetry restoration: The correlation functions for the
Goldstone \PI\ and the non-Goldstone \PI\ look rather different.

\clearpage

\section{Meson Propagators and Spectroscopy} \label{sect-spectroscopy}

\noindent We now compare the correlation functions found by solving the
 order $g_0$ \SDeqs\ with those from \MC. We show some \MC\ propagators
together with the \SD\ correlation functions calculated
with the same bare mass and the coupling chosen such that the renormalised
fermion masses agree.
In view of the fact that the order $g_0$ \SDeqs\
do not reproduce exactly
the critical coupling
found in the \MC\ calculations it is not surprising that a shift in the
coupling is necessary to achieve agreement in the propagators. Indeed the shift
is about the same as the difference between the critical couplings.
Looking at figures~\ref{fig-prop-pi-comp}--\ref{fig-prop-rho-comp}
we find a good agreement between the
\SD\ and \MC\ data. The agreement is of course best at small $g_0$ but
even at large $g_0$ is still satisfactory.

\ifig{b}{fig-prop-pi-comp}{%
Comparison of \SD\ and \MC\ data for the \PI\ propagators $C_2^{(0)}(t)$
(solid lines, circles) and $C_2^{({\rm w})}(t)$ (dashed lines, diamonds).
The results come from calculations on an $8^3 \times 16$ lattice at $m_0 =
0.02$ and $g_0 = 0.27$ (\MC). For the \SD\ calculation we used
$g_0 = 0.2491$ leading to the same
fermion mass as the \MC\ calculation.}

\clearpage

Like the \MC\ measurements
the \SDeqs\ give a rather small annihilation contribution to the
correlation functions. The largest effect is seen for the \SIGMA\
particle deep in the broken phase. An example is shown in
\Fig{fig-anni-sigma-sd}. In distinction to the \MC\ approach there are
no noise problems.
Let us note that the order $g_0$ \SDeqs\ satisfy the Ward
 identity~\cite{Kil87,Pat84}
\begin{equation}
 \frac{d\CHI}{d m_0} = \sum_t (-1)^t C_1^{(0)}(t)\,,
\end{equation}
if one uses the full correlation functions (but not if the annihilation
contribution is neglected).

\ifig{t}{fig-prop-sigma-comp}{%
Comparison of \SD\ and \MC\ data for the $\pi' / \sigma$ propagators.
Symbols and parameters are the same as in \Fig{fig-prop-pi-comp}. Solid
symbols represent negative values.}

{}From now on we just look at the fermion line connected part of the
propagators. In figures~\ref{fig-levels-pi}--\ref{fig-levels-rho}
we compare energy levels from
the \MC\ with those from the \SD\ calculations.
In these figures we
\clearpage
\noindent
 plot some measured values from
tables~\ref{tab-pi}--\ref{tab-rho} against the renormalised fermion
mass.
The $s$-wave states are represented by solid lines (\SD) and solid circles
(\MC), the $p$-wave states by dashed lines (\SD) and open circles (\MC).
The dotted lines represent $2 \mR$, so bound states lie below this
line. The states that lie above the threshold would form the continuum on
the infinite lattice possibly including resonances. One must be
careful about states that lie slightly below the threshold because they
might appear to be bound due to the finite volume.

\ifig{t}{fig-prop-rho-comp}{%
Comparison of \SD\ and \MC\ data for the $\rho / b$ propagators.
Symbols and parameters are the same as in \Fig{fig-prop-pi-comp}. Solid
symbols represent negative values.}

Let us discuss the pictures in more detail. The \PI\ picture
(\Fig{fig-levels-pi})
shows that the
first two levels are in very good agreement.
It is thus reasonable
to expect that the higher levels are correctly represented by the \SD\
formalism.
The large gap between the second and third level explains
 the success of the
two level fit formula.
Obviously we find a bound state with the quantum numbers of the \PI\
in the broken phase.
This state is the pseudo-Goldstone boson \, associated
with chiral symmetry
breaking, and its

\clearpage

\ifig{t}{fig-anni-sigma-sd}{%
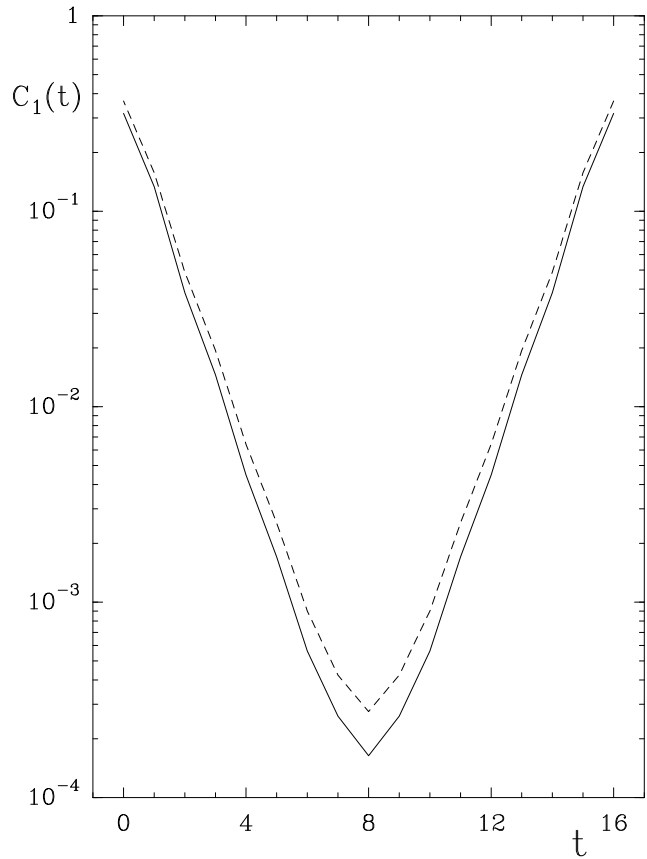
Influence of the annihilation part on the $\pi' / \sigma$ propagator in the
\SD\ calculation. The solid line is $C_1^{(0)}(t)$ calculated without the
annihilation part, the dashed line represents the result including it. The
parameters are the same as in \Fig{fig-anni-sigma-mc}(b), except that $g_0 =
0.2975$ leading to the same fermion mass as the \MC\ data at $g_0 = 0.32$.}

\noindent
 mass tends to zero as $m_0$ tends to zero as one expects.
The picture shows avoided level crossing
suggesting the possible existence of a resonance~\cite{Lus91} in the symmetric
phase.

In the $\pi'/\sigma$ picture (\Fig{fig-levels-sigma})
we see two states with opposite parity
(the \SIGMA\ is the top curve). For small $\mR$ the ground state energies are
reproduced by the \SDeqs. There is somewhat less good agreement for large
values of $\mR$. It is not clear whether we have a slightly bound state for the
\SIGMA. The ground state energy of the $\pi'$ is always close to $2 \mR$.

Finally in \Fig{fig-levels-rho}
we show levels in the $\rho/b$ channel (the \RHO\ is the
solid curve). The agreement for the \RHO\ ground state is good, for the excited
states the fit cannot really resolve the levels lying relatively close to each
other. The ground state of the \RHO\ does not appear to be bound. The $\rho'/a$
and $\omega/b'$ pictures look similar.

\clearpage

\hfig{p}{fig-levels-pi}{%
Energy levels in the \PI\ channel on $8^3 \times 16$ (\MC) and $8^3 \times
\infty$ lattices (\SD) at $m_0 = 0.02$. The dotted line represents the
threshold $E_\pi = 2\mR$.}
\hfig{p}{fig-levels-sigma}{%
Energy levels in the $\pi' / \sigma$ channel on $8^3 \times 16$ (\MC) and $8^3
\times \infty$ lattices (\SD) at $m_0 = 0.02$. The dotted line represents the
threshold $E_{\pi'/\sigma}= 2\mR$. The solid line and the black symbols
represent $E_{\pi'}$ while the dashed lines and open symbols represent
$E_{\sigma}$.}
\clearpage

\PSfigure{t}{(14.1,11.1)}{(0,1.5)}{fig-levels-rho}{fig-levels-rho}{%
Energy levels in the $\rho / b$
channel on $8^3 \times 16$ (\MC) and $8^3 \times
\infty$ lattices (\SD) at $m_0 = 0.02$.
The dotted line represents the
threshold $E_{\rho/b} = 2\mR$. The solid line and the black symbols
represent $E_{\rho}$ while the dashed lines and open symbols represent
$E_{b}$.}

We now look at what is happening in large volumes. In \Fig{fig-large-vol-pi}
we have plotted
the energy levels of the \PI\ on a $52^3 \times \infty$ lattice (solid lines).
Also shown are the ground state
energies for smaller lattices together with \MC\
data. For large $\mR$ we have a bound state which for small values of $\mR$
turns into a resonance, whose energy was calculated from the infinite volume
spectral function (see below). In the finite volume the resonance reveals
itself by avoided level crossing. Finite size effects for the ground state are
largest close to the threshold.

For comparison we also display energy levels of the \SIGMA\ and \RHO\ on
a large lattice along with the resonance energies
(figures~\ref{fig-large-vol-sigma} and~\ref{fig-large-vol-rho}). In both
cases there are bound states for large $\mR$. We would like to point out that
the \RHO\ resonance and bound state energies
stay large even as $m_0 \rightarrow 0$ (see below).

\clearpage

\Hfig{t}{fig-large-vol-pi}{%
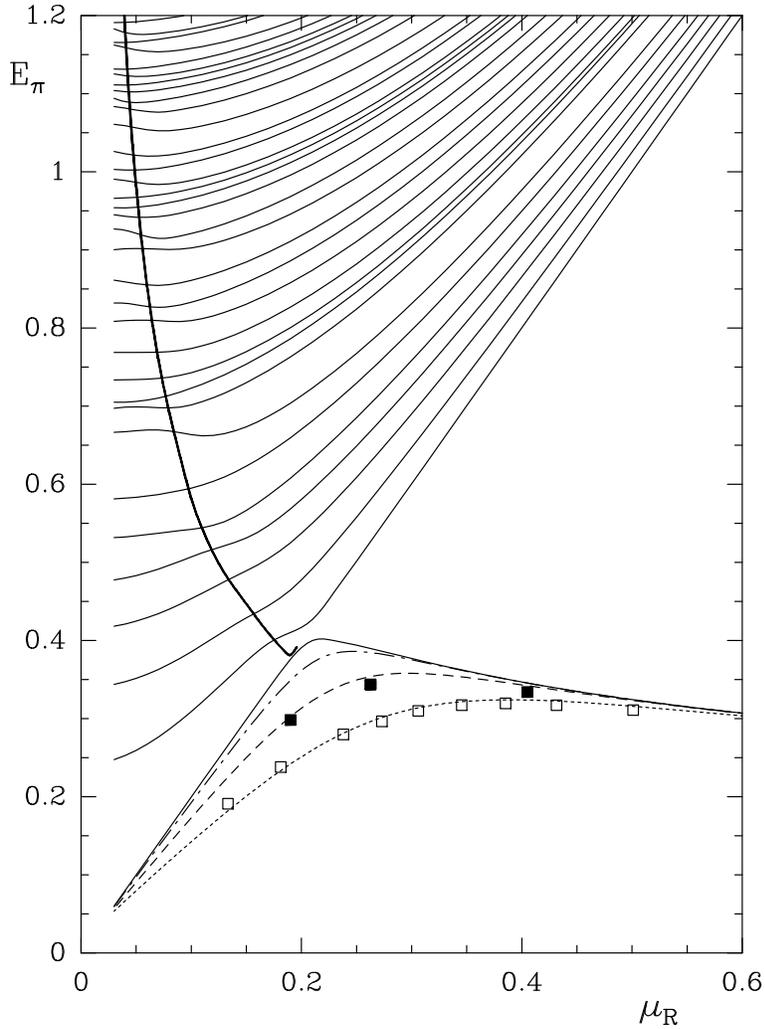
Energy levels of the \PI\ particle on a $52^3 \times \infty$ lattice $(m_0 =
0.02)$. Also shown are the ground state energies on $20^3 \times \infty$,
$12^3 \times \infty$ and $8^3 \times \infty$ lattices together with \MC\
results from $12^4$ and $8^3\times16$ lattices.
The solid line indicates the resonance energy at infinite volume.}

As the volume increases the density of levels above threshold
increases further until for infinite volume we have a continuum and can now
define spectral functions $\rho_{\pm}^{(l)}$, $l \in \{0,1,\rm w\}$:
\begin{equation}
 C^{(l)}(t) = \int\limits_0^{\infty} \frac{dE}{\pi} e^{-E |t|} \rho_{+}^{(l)}(E)
 + (-1)^t \int\limits_0^{\infty} \frac{dE}{\pi} e^{-E |t|} \rho_{-}^{(l)}(E) \,.
\end{equation}
They are calculated from the discontinuities across the cuts and poles
of the amplitude $\tilde{C}$ in \Fig{fig-cuts}:
\begin{eqnarray}
 \rho^i_{+} & = & \Im \tilde{C}^i(iE + \epsilon) \,, \nonumber \\
 \rho^i_{-} & = & \Im \tilde{C}^i(iE + \pi + \epsilon)  \,,
  \ \epsilon \rightarrow +0  \,.
\end{eqnarray}
\clearpage

\Hfig{t}{fig-large-vol-sigma}{%
Energy levels of the \SIGMA\ particle on a $20^3 \times \infty$ lattice ($m_0 =
0.02$). The solid line indicates the resonance energy at infinite
volume. The dotted line is the threshold.}

Let us look at the \PI\ spectral function in more detail. In
this case we need only one spectral function $\rho_{+}$, because
$\rho_{-}$ is zero. Above $2 \mR$ there is a
  continuum of states
(\Fig{fig-spec-pi}b). In the symmetric phase \Fig{fig-spec-pi}a shows
a resonance. We can tell that this `bump' is really a resonance
because we have analytically continued the amplitude to the second
Riemann sheet and found a pole. Locating the pole gives us both the
real and imaginary parts of the pion mass and so tells us the rate
$\Gamma_\pi$ at which the \PI\ resonance decays to a fermion-antifermion
pair. In the broken phase the spectral function of the \PI\ contains a
$\delta$-function contribution. This shows that the resonance has
turned into a bound state.

\clearpage

\Hfig{t}{fig-large-vol-rho}{%
Energy levels of the \RHO\ particle on a $20^3 \times \infty$ lattice ($m_0 =
0.02$). The solid line indicates the resonance energy at infinite
volume.}

In \Fig{fig-width-pi} we show the \PI\ resonance mass $m_\pi$ and width
$\Gamma_\pi$ in  the symmetric phase. Note that the ratio $\Gamma_\pi / m_\pi$
decreases towards zero as we approach the critical point, suggesting that the
physical strength of the meson-fermion coupling decreases as
we go to the continuum limit. This is what we would expect in a
`trivial' theory.

In \Fig{fig-spec-sigma} we show the spectral function of the \SIGMA.
There is a strong and
narrow resonance just above the threshold. This corresponds to the scalar state
at $2 \mR$ that Nambu and Jona-Lasinio found~\cite{Nam61}.

\PSfigure{p}{(14.1,18.0)}{(0,0.6)}{fig-spec-pi}{fig-spec-pi}{%
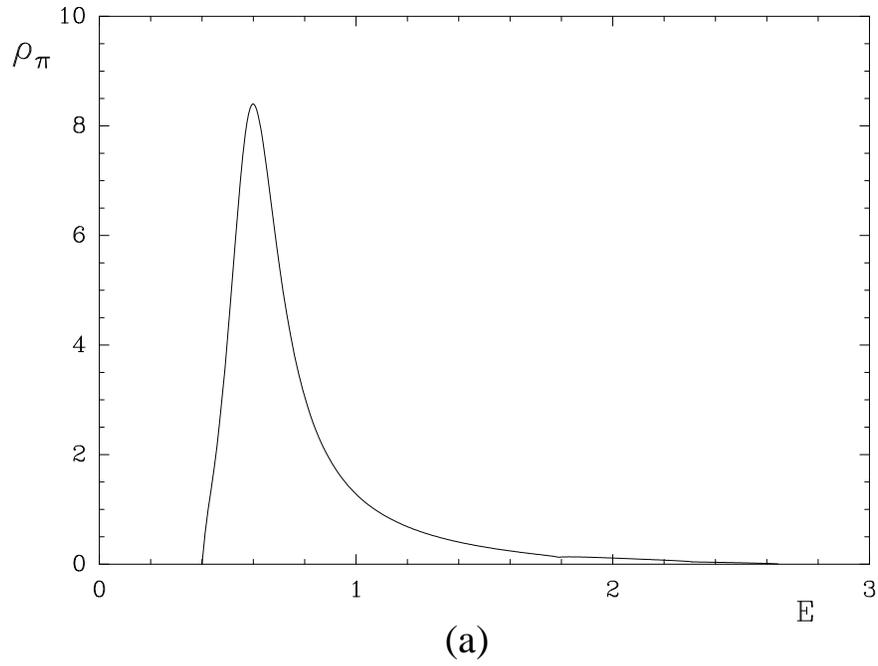
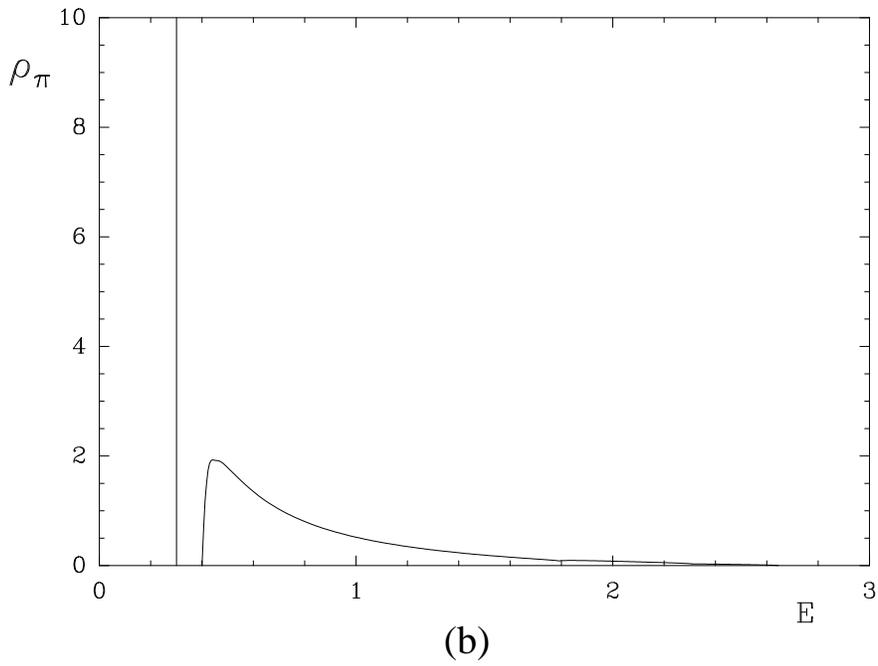
Spectral function of the \PI\ particle:
\aaa\ in the symmetric phase ($m_0=0.04$, $g_0=0.2056$
leading to $\mu_R=0.20$),
\bbb\ near $g_c$ ($m_0=0.01$, $g_0=0.2476$ leading to $\mu_R=0.20$).}

\hfig{p}{fig-width-pi}{%
Resonance mass (solid line) and width (dotted line) of the \PI\ particle
at $m_0=0$.}

\hfig{p}{fig-spec-sigma}{%
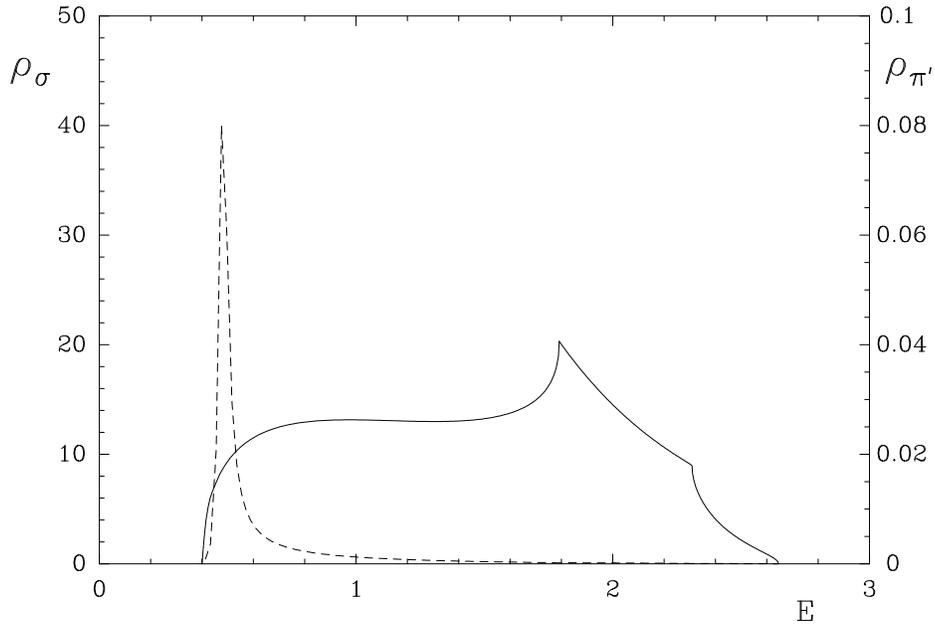
Spectral function of the \SIGMA\ (dashed line) and $\pi'$ (solid line) particles
 near $g_c$ ($m_0=0.01$, $g_0=0.2476$ leading to $\mu_R=0.20$).
(Note that the two functions have been shown at very different scales.)}

\clearpage

\hfig{t}{fig-spec-rho}{%
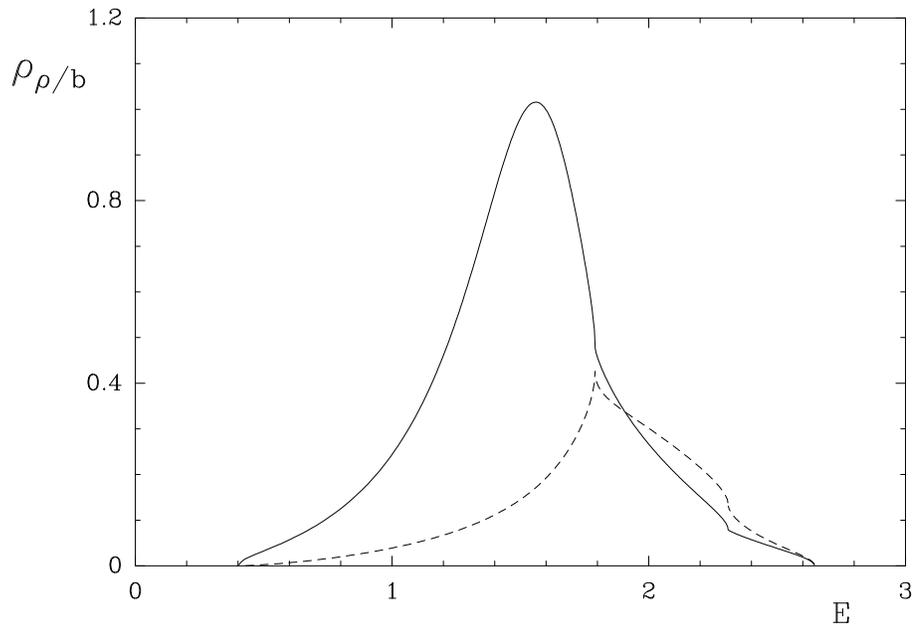
Spectral function of the \RHO\ (solid line) and $b$ (dashed line) particles
in the symmetric phase ($m_0=0.04$, $g_0=0.2056$, leading to $\mu_R=0.20$).}

\hfig{b}{fig-rho-width}{%
Resonance mass (solid line) and width (dotted line) of the $\rho$ particle
at $m_0 = 0.02$.}

Next we look at the picture of the \RHO\ and $b$ spectral functions
(\Fig{fig-spec-rho}). We see that the \RHO\ is a resonance. Although
this would initially appear interesting
  we find that this resonance
never becomes light compared with the
inverse lattice \, spacing. Even at
the critical \, point it has a mass of \, about 1.5 (see

\clearpage

\noindent \Fig{fig-rho-width}).
In the $b$ channel
there is only a continuum. Note
that the $s$\/- and
   $p$\/-wave show a different threshold behaviour
($\rho \propto [E - 2\mR]^{1/2}$ for the $s$\/-wave,
 $\rho \propto [E - 2\mR]^{3/2}$ for the $p$\/-wave).

  As well as the resonance we see two other curious features,
 sharp discontinuities in the slope of the spectral function
 at energies of
 approximately 1.8 and 2.3. In general they appear at
  $E  =  \cosh^{-1}(3 + 2 \sinh^2 \mR)$ and
  $ \cosh^{-1}(5 + 2 \sinh^2 \mR)$.
  These are examples of  van Hove
 singularities~\cite{vanHove}. Such singularities are
 familiar from solid state physics and result from saddle points in the
 fermions'
 energy momentum relation. At the corresponding
 energies the density of states is singular leading to singularities
 in the spectral functions (see appendix \ref{app-inf-vol}).
 These singularities are also present (though often less
 noticeable) in the other channels. The van Hove singularities
 introduce a complication when analytically continuing onto the
  non-physical Riemann sheet (see \Fig{fig-cuts}).
  Because there are branch
  points we reach different sheets depending on
  where we cross the axis in relation to these singularities. The
  sheet which is relevant for the continuum limit of the theory is
  the one reached by crossing the axis between the threshold and
  the first van Hove singularity. The other sheets only exist
  because of the saddle points in the lattice fermion dispersion
  relation at $\vec{k}=(\frac{\pi}{2},0,0),
  (\frac{\pi}{2}, \frac{\pi}{2},0)$
  etc., and so should have no relevance to the continuum. All the
  resonances we have seen are on this `low energy' sheet.

The $\rho'/a$ channel is similar to the $\rho / b$. We see
similar resonances (one of each parity) which in the
continuum limit go to the cutoff. For the $\omega / b'$ the \MC\ results
indicate that these particles are also heavy.

\clearpage

\section{Renormalisation Group Flow}

\noindent In this section we study the renormalisation group flow in our
 model. We consider dimensionless ratios of physical quantities and attempt
 to find lines of constant physics, i.~e., lines where all those ratios
are constant, independent of the cutoff. Only in regions where such
 lines exist is the theory renormalisable~\cite{Goc92,Lepage,Brown}.
 Since we are interested in physical quantities
 we should only look at the ratios involving
 the particles whose correlation lengths diverge at the critical
point, namely the fermion, \PI\ and \SIGMA. Other states (such as the
 \RHO) with energies near $1/a$ are irrelevant in this context.

\hfig{b}{fig-flow-mc}{%
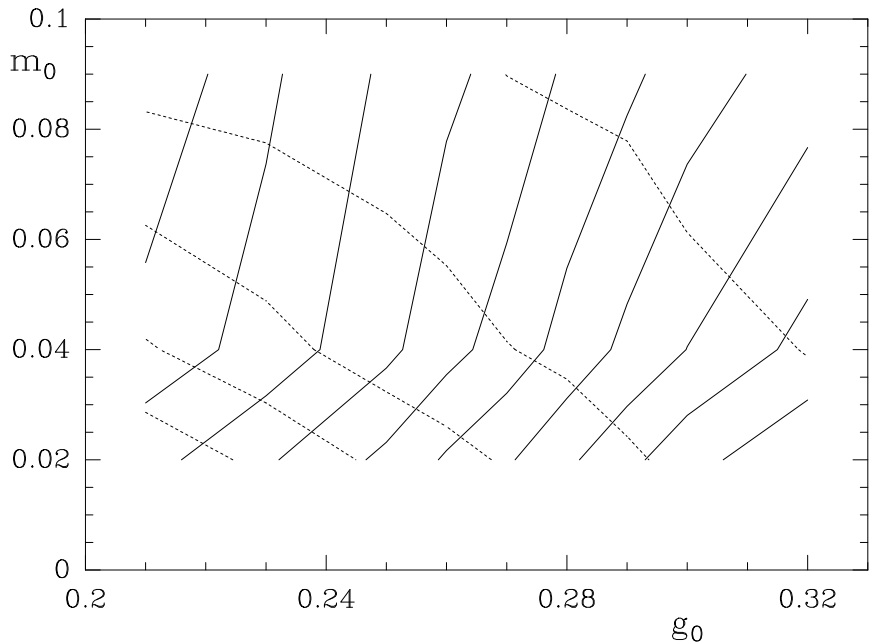
Lines of constant $\Epi / \mR$ (solid) and of constant $\mR / \fpi$ (dotted)
in the plane of the bare parameters from \MC\ results on an $8^3 \times 16$
lattice. $\Epi / \mR = 0.7$ to $1.6$ in
 steps of $0.1$ (right to
left). $\mR / \fpi = 1.5$ to $3.5$ in steps of $0.5$ (left to right).}

In \Fig{fig-flow-mc} we show $\Epi / \mR$ and $\mR / \fpi$ for the
\MC\ results on an $8^3 \times 16$ lattice. There is no region where
the lines are parallel to each other, thus there are no lines of
constant physics. There are however two caveats to this picture. The
first is that we have large \FS\ effects and secondly that in the
symmetric phase the \PI\ is a resonance, and so $\Epi$ no
longer corresponds to the pion mass (see \Fig{fig-large-vol-pi}).
The upper right hand corner
suffers least from these problems, and here the Monte Carlo results
 can be taken at face value.
 These difficulties can be circumvented by
use of
 \SDeqs. In \Fig{fig-flow-pi-sd} we show the \SD\ results on
$8^3 \times \infty$ and $20^3 \times \infty$ lattices.
 Comparison of the two
volumes shows the
  features which are robust. In the broken phase the
lines of constant mass ratio flow towards the critical point but curve
away \, just before reaching it.

\PSfigure{p}{(14.1,18.0)}{(0,0.6)}{fig-flow-pi-sd}{fig-flow-sd}{%
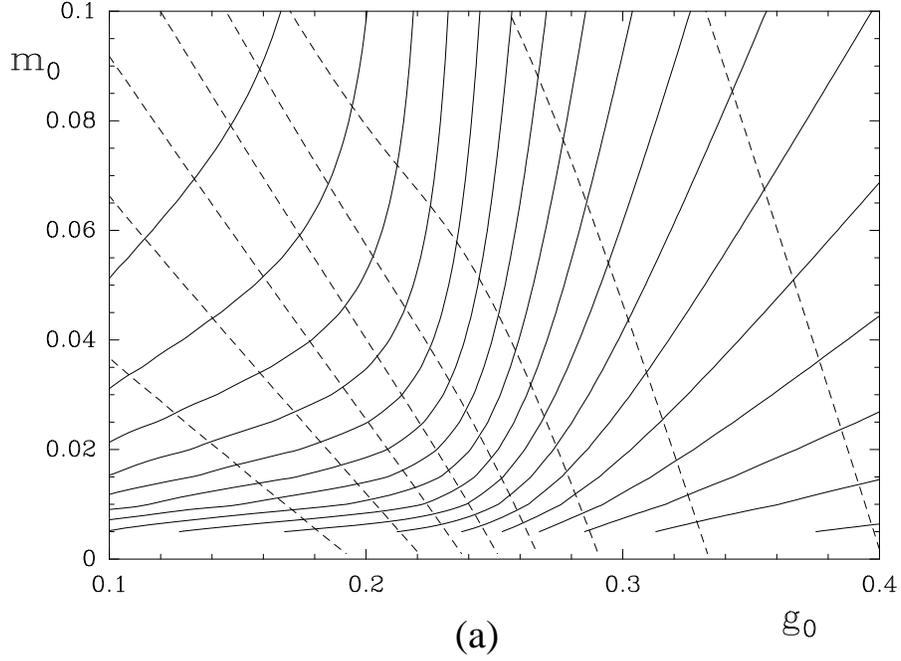
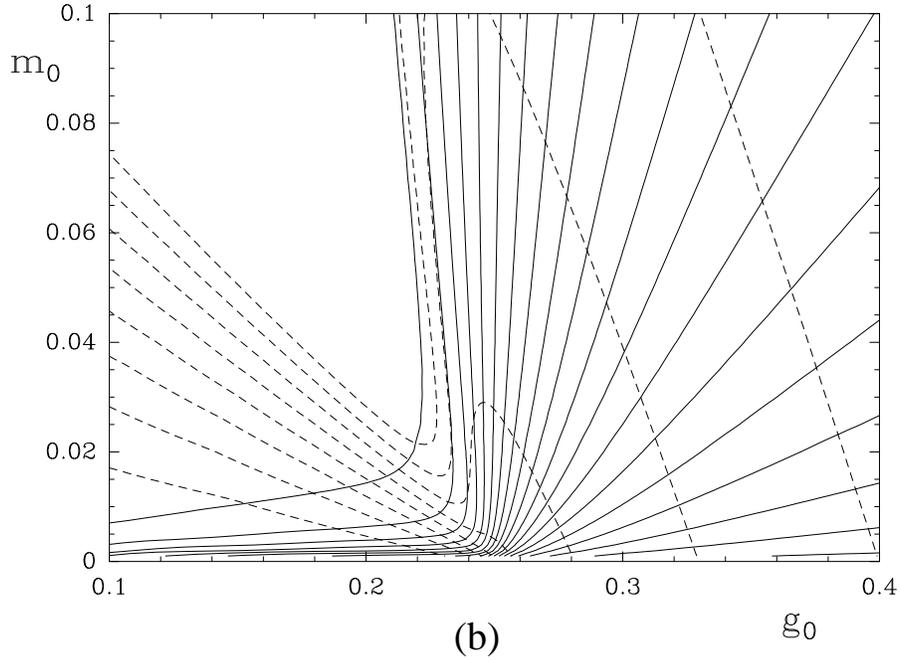
Lines of constant $\Epi / \mR$ (solid) and
of constant $\mR / \fpi$ (dashed) from \SD\ calculations
 on $8^3 \times \infty$
\aaa\ and $20^3 \times \infty$ \bbb\ lattices.
For $\Epi / \mR$ the ratios are:
\aaa\ $0.2$ to $1.8$ and \bbb\ $0.1$ to $1.9$ in steps of $0.1$.
In both figures $\mR / \fpi = 0.5$ to $4.0$ in steps of $0.5$.}

\clearpage

\noindent In the symmetric phase the ratio is
always just below 2 because we have used $\Epi$, the lowest energy
level in the \PI\ channel.
 (When comparing these figures with \Fig{fig-flow-mc} remember that
 the equivalence between the \MC\ and the \SDeqs\ is not exact,
 there being a small shift in $g_c$ between them.)

Extrapolating to infinite volume by the \SDeqs\ gives \Fig{fig-flow-pi-inf}.
In this limit we can identify resonances and find their masses.
 These are shown in the
 diagram as dotted lines. Now all the lines flow into the critical
point. This type of flow diagram was also found in QED~\cite{Goc92}.
In the broken phase we can see that the lines of constant mass
 ratio, \Fig{fig-flow-pi-inf}, and the lines of constant
 $\mR / \fpi$, \Fig{fig-flow-fpi-inf}, take different paths
 just as in \Fig{fig-flow-mc},
 again showing that there are no lines of constant physics.

   To test for the existence of lines of constant physics in the
 symmetric phase we have looked at the ratio between $m_\pi$
 and $\Gamma_\pi$. These quantities are found from
 the real and imaginary parts of the pole position. In \Fig{fig-flow-gpi}
 lines of constant $\Gamma_\pi / m_\pi$ are compared with curves of constant
 $\mR / m_\pi$. Once again the different flows cross. In this phase too
 there are no lines of constant physics. To display this information
 another way we show in \Fig{fig-gpi} the way in which
 $\Gamma_\pi / m_\pi$ varies along a particular curve of constant
 mass ratio ($\mR / m_\pi = 0.20$). Because the phase space available
 for the decay depends on the mass ratio the variation in
 $\Gamma_\pi / m_\pi$ along this path is not due to kinematics, but
 must \, reflect a variation
 in the strength of the coupling \, between the
 \PI\

\hfig{b}{fig-flow-pi-inf}{%
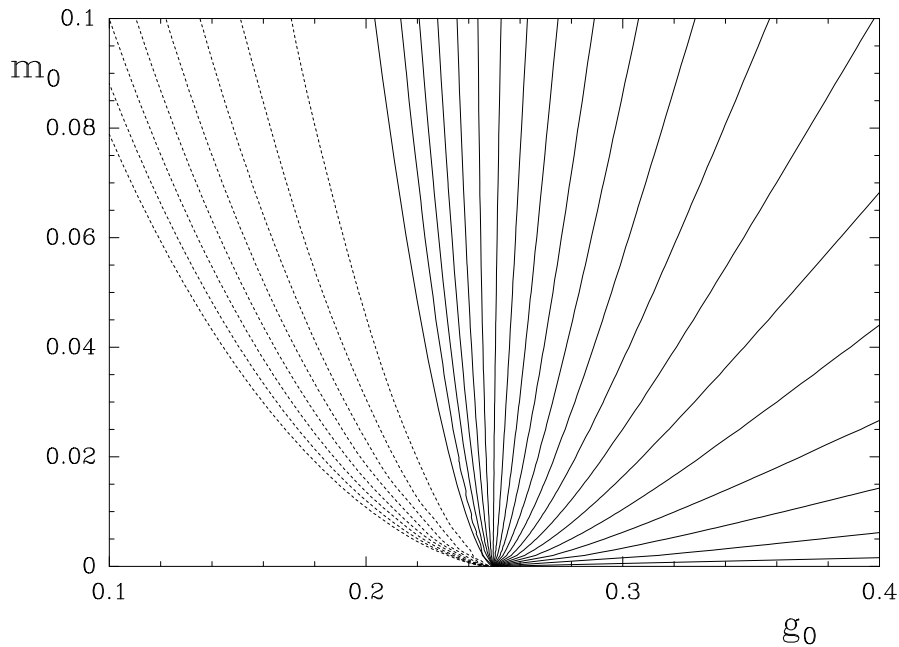
Lines of constant $m_\pi / \mR$ on an $\infty^4$ lattice.
Solid lines are used where the
\PI\ is a bound state and dotted lines where it is a resonance. The ratios are
$0.1$ to $2.0$ in steps of $0.1$ (bound state)
 and $3$ to $10$ in steps of $1$ (resonance).}

\clearpage

\hfig{p}{fig-flow-fpi-inf}{%
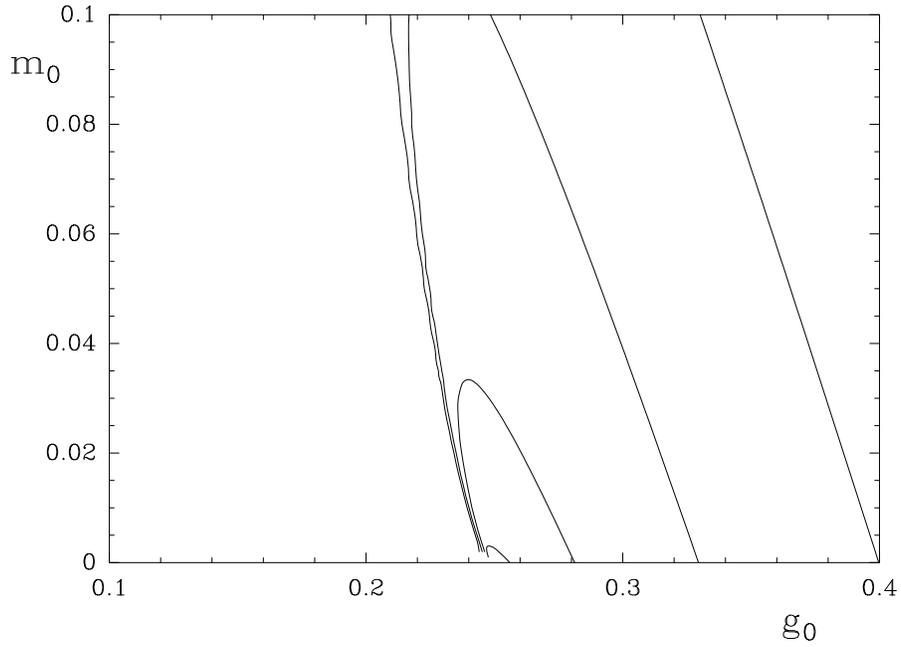
Lines of constant $\mR / \fpi$ on an $\infty^4$ lattice.
The ratio runs from $2.5$ (inner curve) to $4.0$ (outer curve) in
steps of $0.5$.}

\hfig{p}{fig-flow-gpi}{%
 A comparison between the flows of constant $\Gamma_\pi / m_\pi$
 (solid lines) and constant $\mR /m_\pi$ (dotted lines) in
 the region where the \PI\ is a resonance.
 $\mR / m_\pi$ runs from $0.1$ to $0.4$ in steps of $0.1$,
 $\Gamma_\pi /m_\pi$ from $0.1$ to $0.3$ in steps of $0.05$. }

\hfig{t}{fig-gpi}{%
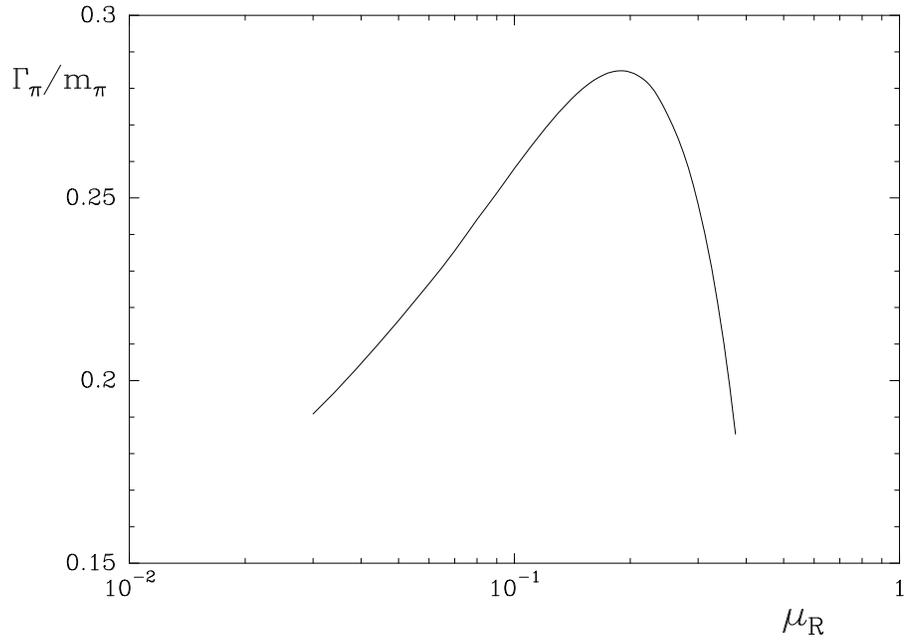
 The variation in $\Gamma_\pi / m_\pi$ along the curve
 $\mR /m_\pi = 0.20$.}

\hfig{b}{fig-gsigma}{%
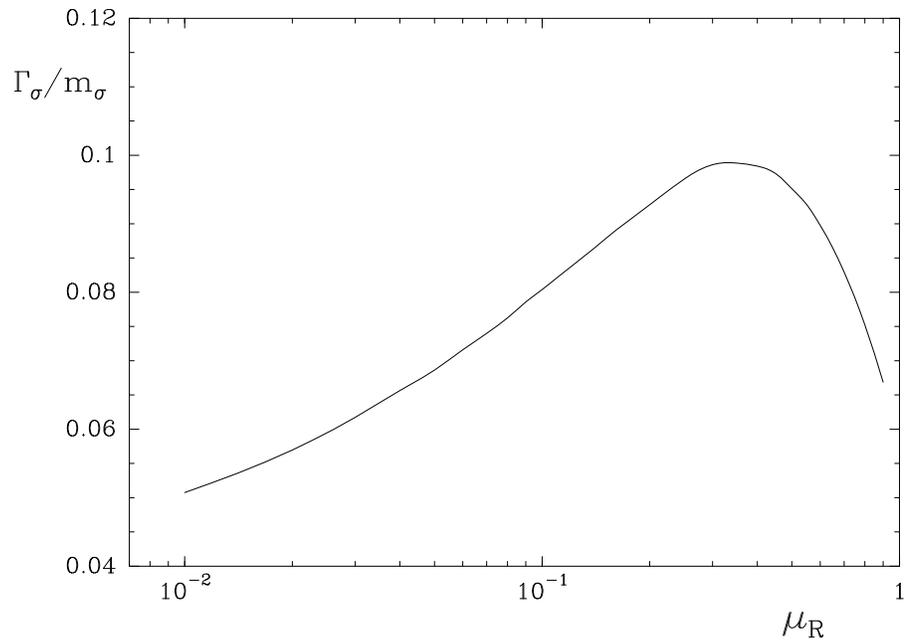
 The variation in $\Gamma_\sigma / m_\sigma$ along the curve
 $\mR /m_\pi = 1/2$.}

\hfig{p}{fig-flow-sigma-inf}{%
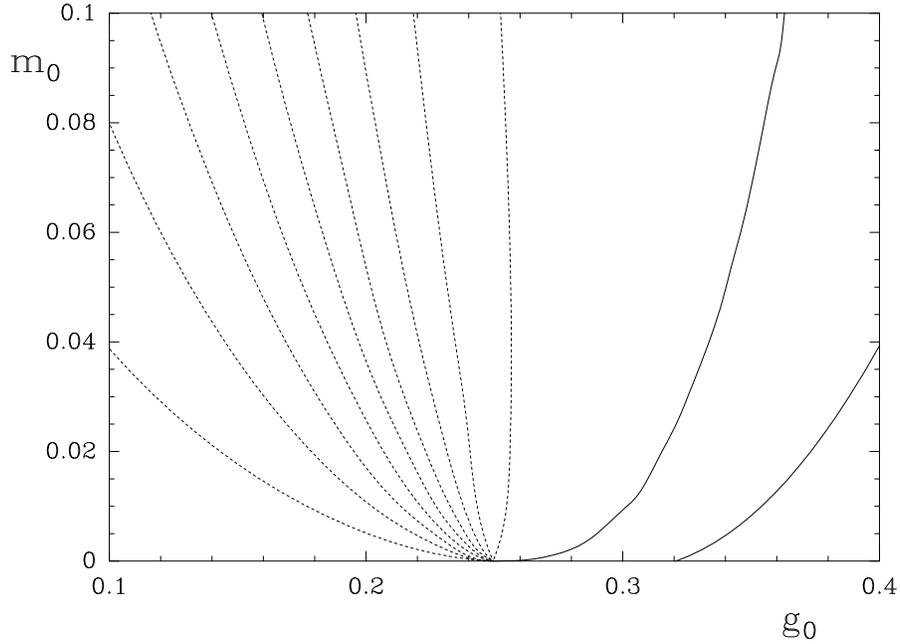
Lines of constant $\mR / m_\sigma$ on an $\infty^4$ lattice.
The ratios are $0.51$ and $0.50$ (bound state and threshold: solid lines);
$0.45$ to $0.05$ in steps of $0.05$ (resonance: dashed lines).}

\hfig{p}{fig-flow-gsigma}{%
 A comparison between the flows of constant $\Gamma_\sigma / m_\sigma$
 (solid lines) and constant $\mR /m_\sigma$ (dotted lines) in
 the region where the \SIGMA\ is a resonance.
 $\mR / m_\sigma$ runs from $0.05$ to $0.45$ in steps of $0.05$,
 $\Gamma_\sigma /m_\sigma$ from $0.0$ to $0.3$ in steps of $0.05$. }

\clearpage

 \noindent
 and the fermion. As we approach the critical point, $\mR$ small,
 this physical coupling decreases, consistent with the
 hypothesis that the NJL model is trivial.

     Note that on the curve $m_\pi = 2 \mR$, the threshold for pion decay,
  both $\Gamma_\pi$ and $f_\pi$ vanish. This raises the
 possibility that this  threshold could be a line of constant physics.
 To check this we have to look at other particles such as the \SIGMA.
 In \Fig{fig-gsigma} we plot the variation of the ratio
 $\Gamma_\sigma / m_\sigma$ along the line $m_\pi = 2 \mR$. We see that the
  decay rate of the \SIGMA\ is not constant along this curve, but decreases
  slowly as we approach the critical point. Fig.~\ref{fig-gsigma} is in fact
  reminiscent of \Fig{fig-gpi}, and again consistent with a trivial
  continuum limit for the NJL model.

 In \Fig{fig-flow-sigma-inf} we show the ratio of fermion to \SIGMA\
 mass as found from the \SDeqs. Above $g_c$ there is a large region
 where this ratio is always very
 close to the value of $0.5$ found in~\cite{Nam61}. In the symmetric phase
 the \SIGMA\ is a resonance with a mass similar to that of the \PI, as
 would be expected from unbroken chiral symmetry.
 (Because the \SIGMA\
 mass hovers around $2 \mR$ in the broken phase the mass ratio
 $E_\sigma^{(0)} / \mR$ is
 essentially constant and a Monte Carlo flow picture shows nothing new.)
 In \Fig{fig-flow-gsigma} we compare the flow patterns for the ratios
 $\Gamma_\sigma / m_\sigma$ and $\mR / m_\sigma$. As in \Fig{fig-flow-gpi},
 the corresponding picture for the pion, we see a crossing of the two flows,
 once again a sign of non-renormalisability.

\clearpage

\section{Conclusions}

We have made a thorough investigation of a lattice version of the NJL
model using both the \MC\ method and \SDeqs. The interplay between
both methods allowed us to come much further than with either method
alone. We extensively used the \SDeqs\ because of rather large \FS\
effects.

Since we were interested in the chiral symmetry properties we worked
with staggered fermions. A phase transition was seen at about $g_c
\approx 0.280(4)$. In the Goldstone \PI\ channel we have successfully
identified a bound state in the broken phase and a resonance in the
symmetric phase. In the \SIGMA\ and \RHO\ channels we have seen
resonances. We have found that particles in the \PI\ and \SIGMA\
channels become massless at the critical point while particles in the
\RHO\ channel scale with the inverse lattice spacing.

Equipped with the results from spectroscopy we worked out
renormalisation group flows in the bare parameter plane. We do not
find any lines of constant physics: lines of constant $\mR / \fpi$
cross lines of constant $m_\pi / \mR$. The ratio $m_\sigma / \mR$ has
come out essentially constant in the broken phase
 and so has no bearing on this question.
The absence of renormalisability implies that one should be cautious
in applying renormalisation group techniques as is sometimes done in
the top-mode standard model.

 The NJL model has attracted speculations such
as a non-Lorentz invariant vacuum or a massless vector
 state~\cite{others}. These do not appear to be realised at
the phase transition that we investigated.

Our lattice model can of course be embedded in a generalized NJL model
with more bare parameters~\cite{Higgs}. It might then be possible to
find regions in this extended parameter space where the model is
(weakly) renormalisable.
If such a region intersects the part of the parameter space we have studied our
physical results (absence of light vector states, occurrence of resonances,
etc.) would hold inside this whole region.

We have now reached a comprehensive understanding of the four dimensional
NJL-Model concerning the chiral condensate, the fermion mass, meson
spectroscopy and renormalisation properties.

\vspace{0.5cm}

\section*{Acknowledgements}

This work was in part supported by the Deutsche Forschungsgemeinschaft.
The numerical calculations were performed on
the Fujitsu VP 2400/40 at the RRZN Hannover,
the Crays at the ZIB Berlin and the Cray X-MP at the RZ Kiel.
We thank all these institutions for their support.

\clearpage

\appendix

\section{\SD\ Equations for the Composite States} \label{app-SDeqs}

In this appendix we shall derive meson correlation functions using the
\SDeqs\ shown in \Fig{fig-sd-c}. $K$ is the two particle irreducible
kernel, which is in general a function of 4 coordinates, and so on a
lattice of volume $V$ has $V^4$ values. This is reduced by a factor
$V$ due to translation invariance. If we consider a
fermion-antifermion pair with definite centre of mass momentum $p$,
momentum conservation further reduces this to $V^2$ values. $K$ can
therefore be
regarded as a $V \times V$ matrix that links the separation vector of the
incoming fermion-antifermion pair (with $V$ possible values) to
the relative distance of the outgoing pair. This becomes much more
tractable when we replace the full kernel by the bare four fermion
interaction. The resulting equations are at the same level of accuracy
as the order $g_0$ \SDeqs\
which were used for the fermion propagator and
the chiral condensate.

Because the bare kernel is very short range, involving only fields
separated by distance 0 or 1, we only need the two particle wave
function at 9 values of the separation (0 or $\pm \hat{\mu}$). This
allows us to write the \SDeqs\ in a $9 \times 9$ matrix form, so we
can express the solution of them in terms of the momentum space
meson propagators
\begin{equation}
 P_{ij}(p) \equiv \frac{1}{V} \langle {\cal O}_i(-p) {\cal O}_j(p) \rangle \,,
\vspace{-0.2cm}
\end{equation}
where
\begin{eqnarray}
\vspace{-0.1cm}
 {\cal O}_0(p) &=& \sum_x \bar{\chi}(x) \chi(x) \exp (i p\cdot x)\,,\nonumber\\
\vspace{-0.1cm}
 {\cal O}_j(p) &=& \sum_x \eta_{\mu}(x) \bar{\chi}(x) \chi(x + \hat{\mu})
      \exp (i p\cdot (x + \frac{1}{2} \hat{\mu} )), \; j = 2 \mu \,,\nonumber\\
\vspace{-0.1cm}
 {\cal O}_j(p) &=& \sum_x \eta_{\mu}(x) \bar{\chi}(x+\hat{\mu}) \chi(x)
      \exp (i p\cdot (x + \frac{1}{2} \hat{\mu} )), \; j = 2 \mu - 1 \,,
\end{eqnarray}
and $\eta_{\mu}(x)$ is defined in (\ref{eta-defn}).
In this basis the four fermion kernel $K$ is given by
%\clearpage
%
\begin{equation}
K = \left(
\begin{array}{|c|cc|cc|cc|cc|}
\cline{1-1}
-2 g_0 \sum_\mu \cos p_\mu  &       \multicolumn{8}{c}{} \\
\cline{1-3}
\multicolumn{1}{c|}{} &  0  & g_0 & \multicolumn{6}{c}{} \\
\multicolumn{1}{c|}{} & g_0 &  0  & \multicolumn{6}{c}{} \\
\cline{2-5}
\multicolumn{3}{c|}{} &  0  & g_0 & \multicolumn{4}{c}{} \\
\multicolumn{3}{c|}{} & g_0 &  0  & \multicolumn{4}{c}{} \\
\cline{4-7}
\multicolumn{5}{c|}{} &  0  & g_0 & \multicolumn{2}{c}{} \\
\multicolumn{5}{c|}{} & g_0 &  0  & \multicolumn{2}{c}{} \\
\cline{6-9}
\multicolumn{7}{c|}{} &  0  & g_0 \\
\multicolumn{7}{c|}{} & g_0 &  0  \\
\cline{8-9}
\end{array} \right) \,, \label{eq-K}
\end{equation}
\clearpage
\noindent
and the \SD\ equation reads
\begin{equation}
 P = M + MKP \,. \label{eq-P}
\end{equation}
 As can be seen from the figure $M$ is given by the independent
propagation of a fermion and antifermion, i.~e., the $t$-channel
disconnected part of $P$
\begin{equation}
 M_{ij}(p) \equiv \frac{1}{V} \langle {\cal O}_i(-p) {\cal O}_j(p)
   \rangle_{t-{\rm channel\ disconnected }} \, .
\end{equation}
Explicitly
\begin{equation}
\newcommand{\ml}{\multicolumn{1}{|r}}
\newcommand{\mr}{\multicolumn{1}{r|}}
M = \left(
\begin{array}{r|rrrrrrrr}
\cline{1-1}
\multicolumn{1}{|c|}{a}
     &  d_1   & -d_1   &  d_2   & -d_2   &  d_3   & -d_3   &  d_4   & -d_4 \\
\hline
 d_1 &\ml{c_1}&\mr{b_1}& e_{12} &-e_{12} & e_{13} &-e_{13} & e_{14} &-e_{14}\\
-d_1 &\ml{b_1}&\mr{c_1}&-e_{12} & e_{12} &-e_{13} & e_{13} &-e_{14} & e_{14}\\
\cline{2-5}
 d_2 & e_{12} &-e_{12} &\ml{c_2}&\mr{b_2}& e_{23} &-e_{23} & e_{24} &-e_{24}\\
-d_2 &-e_{12} & e_{12} &\ml{b_2}&\mr{c_2}&-e_{23} & e_{23} &-e_{24} & e_{24}\\
\cline{4-7}
 d_3 & e_{13} &-e_{13} & e_{23} &-e_{23} &\ml{c_3}&\mr{b_3}& e_{34} &-e_{34}\\
-d_3 &-e_{13} & e_{13} &-e_{23} & e_{23} &\ml{b_3}&\mr{c_3}&-e_{34} & e_{34}\\
\cline{6-9}
 d_4 & e_{14} &-e_{14} & e_{24} &-e_{24} & e_{34} &-e_{34}&\ml{c_4}&\mr{b_4}\\
-d_4 &-e_{14} & e_{14} &-e_{24} & e_{24} &-e_{34} & e_{34}&\ml{b_4}&\mr{c_4}\\
\cline{8-9}
\end{array} \right) \,. \label{eq-M}
\end{equation}

The ``bubbles" (see \Fig{fig-bubbles}) in $M$ are
\begin{eqnarray}
   a(p) &=&  \frac{1}{V} \sum_k
             \frac{N^2 -\sum_\mu F_\mu^2 s_\mu^{+} s_\mu^{-} }
              {U^{+} \ \ \    U^{-} } \,, \nonumber \\
   b_\mu(p) &=& - \frac{1}{V} \sum_k
             \frac{ F_\mu^2 s_\mu^{+} s_\mu^{-} -
             \sum_{\nu \neq \mu}   F_\nu^2 s_\nu^{+} s_\nu^{-} -N^2 }
             {U^{+} \ \ \    U^{-} }  \,, \nonumber \\
   c_\mu(p) &=& \frac{1}{V} \sum_k
             \frac { \left( F_\mu^2 s_\mu^{+} s_\mu^{-} -
             \sum_{ \nu \neq \mu }
             F_\nu^2 s_\nu^{+} s_\nu^{-} -N^2 \right)
             \left (1 - 2 (s_\mu^0)^2 \right) }{ U^{+} \ \ \    U^{-} }
             \,, \nonumber \\
   d_\mu(p) &=&  \frac{1}{V} \sum_k
             \frac{N F_\mu (s_\mu^{+}+s_\mu^{-}) s_\mu^0 }{U^{+} \ \ \ U^{-} }
             \,, \nonumber \\
   e_{\mu\nu}(p) &=& - \frac{1}{V} \sum_k
             \frac{F_\mu F_\nu (s_\mu^{+}s_\nu^{-}+s_\mu^{-}s_\nu^{+})
             (c_\mu^0 c_\nu^0 +s_\mu^0 s_\nu^0) }{U^{+} \ \ \    U^{-} } \,,
 \label{eq-M-bubbles}
\end{eqnarray}
where
%\clearpage
%
\begin{eqnarray}
   s_\mu^\pm &\equiv& \sin (k_\mu \pm \frac{1}{2} p_\mu), \nonumber \\
   s_\mu^0 &\equiv& \sin k_\mu,  \nonumber \\
   c_\mu^0 &\equiv& \cos k_\mu,  \nonumber \\
   U^{\pm} &\equiv& N^2 - \sum_\mu F_\mu^2 (s_\mu^\pm)^2,
\end{eqnarray}
\clearpage
\noindent
and $N$, $F_\mu$ are given in (\ref{OGfermi}).
Equation~(\ref{eq-P}) is solved by
\begin{equation}
 P = ({\bf 1} - MK)^{-1} M \,.
\end{equation}
This is the full solution including the annihilation term.
 The matrix inversion
is done numerically.

\PSfigure{t}{(14.1,9)}{(0,1)}{fig-bubbles}{fig-bubbles}{%
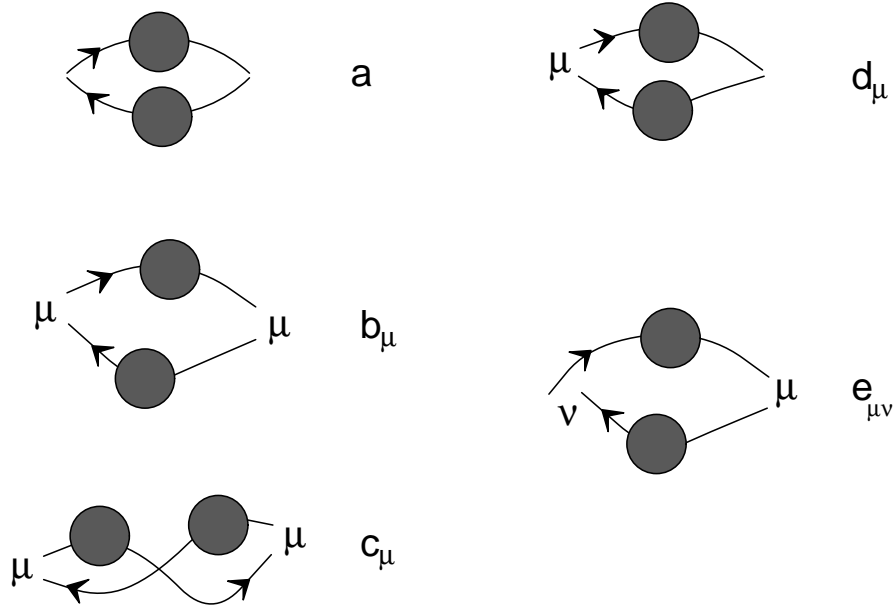
Graphical representation of the ``bubbles'' in eqs.~(\ref{eq-M})
and~(\ref{eq-M-bubbles}).}

     Because we are working with staggered fermions
  we get mesons at rest if the spatial components of
  $p$ are all either $0$ or $\pi$. Other values of $\vec{p}$
  give moving mesons and can be used to check the restoration
  of Lorentz invariance.

    In most of our \MC\ calculations we have measured only
  the fermion line connected part of the meson propagator as is
  customary. This means that t-channel exchange of auxiliary
  fields is kept, but s-channel (i.~e., annihilation) auxiliary fields
  are dropped (see \Eq{eq-picture}). When we make this same approximation in
  the \SDeqs\ only the first term in the kernel is kept.
  We no longer have to consider bond operators,
  so the $9 \times 9$ matrix
   becomes a $1\times 1$ matrix. The meson propagator simplifies to
\begin{eqnarray}
     P(p)& \equiv &
  \sum_x
  \langle (\bar{\chi} \chi)(x) (\bar{\chi} \chi) (0)
 \rangle e^{-i p\cdot x}   \nonumber \\
 & = &\frac{a(p)}{1 + a(p) 2 g_0 \sum_\mu \cos p_\mu }.
 \label{pt-pt}
\end{eqnarray}
\clearpage
\noindent
   The propagator $P(p)$ includes the first four multiplets   shown in
\Tab{tab-names} because the $s$-factors in the table   are all of the
form $\exp(i p \cdot x)$, i.~e., all represent   a simple shift in
momentum in \Eq{pt-pt}.    We are most interested in the propagators
$C_i^{(0)}(t)$ for   stationary mesons, i.~e., mesons for which
$\vec{p}= 0 \mod \pi$.   Let the Fourier transform of $C_i^{(0)}(t)$
be $\tilde{C}_i^{(0)}(\omega)$ (see \Eq{eq-fourier} for normalisation).
The channel   $i=1$ (the $\pi' / \sigma$ channel) corresponds to
\begin{equation}
   \tilde{C}_1^{(0)}(\omega) = P(0,0,0,\pi+\omega) \,,
\end{equation}
$i=2$ (the \PI\ channel) corresponds to
\begin{equation}
   \tilde{C}_2^{(0)}(\omega) = P(\pi,\pi,\pi,\pi+\omega),
\end{equation}
and the $\rho' / a$ and $\rho / b$ channels ($i = 3, 4$) to
\begin{equation}
\tilde{C}_3^{(0)}(\omega) = P(\pi,0,0,\pi+\omega)
                          = P(0,\pi,0,\pi+\omega)
                          = P(0,0,\pi,\pi+\omega)
\end{equation}
and
\begin{equation}
\tilde{C}_4^{(0)}(\omega) = P(0,\pi,\pi,\pi+\omega)
                          = P(\pi,0,\pi,\pi+\omega)
                          = P(\pi,\pi,0,\pi+\omega)
\end{equation}
respectively.
Representing the $\tilde{C}$'s as particular values of $P(p)$
shows that the fact that the different
$i$ and $k$ channels don't mix is simply a consequence of the
conservation of 3-momentum.
Inserting the above momentum values in (\ref{pt-pt}) gives
the results stated in \Eq{eq-C-tilde}.

\clearpage

\section{The Infinite Volume Limit of the Meson Propagators} \label{app-inf-vol}

In this appendix we study the infinite volume limit of the meson
propagators (\ref{eq-C-tilde}).
We have already taken the $L_t \to \infty$ limit in (\ref{calI}).
We now let $L_s \to \infty$.
In the infinite volume limit the different boundary conditions
in  the space and time directions no longer matter and so
$F_s = F_t \equiv F$.

The fermion loop integrals in (\ref{eq-ai}) are all
of the form
\begin{equation}
\int_0^{2\pi} \frac{d k_1}{2\pi} \frac{d k_2}{2\pi} \frac{d k_3}{2\pi}
   h(\sin^2 k_1 + \sin^2 k_2 +\sin^2 k_3) \,. \label{3dint}
\end{equation}
   In the continuum we would naturally simplify such a
 `spherically symmetric' integral by changing variables to
  $k_1^2 + k_2^2 + k_3^2$.  In the same way we use the
  variable
\begin{equation}
    r = -3 + 2 (\sin^2 k_1 + \sin^2 k_2 +\sin^2 k_3)
\end{equation}
  to simplify the lattice integrals. (The shift and normalisation
  make some of the following expressions simpler than they would
  otherwise be.)
The integral (\ref{3dint}) then becomes
\begin{equation}
   \int_{-3}^{3} dr\, h\left(\frac{3}{2}+\frac{1}{2}r \right) \phi(r) \,.
\end{equation}
The `phase space' function $\phi$ is
\begin{eqnarray}
     \phi(r) &=& \int_0^{2 \pi} \frac{d^3 k}{(2 \pi)^3}
      \delta ( r + 3 - 2 (\sin^2 k_1 + \sin^2 k_2 +\sin^2 k_3) )
  \nonumber \\
    &=& \int_{-\infty}^{+\infty} \frac{d \nu} {2 \pi}
             e^{i \nu r} J_0^3(\nu).
\end{eqnarray}
    $\phi$ is 0 outside the range $(-3,3)$ and has branch points
  at $r = \pm 1$. These branch points
  are relevant to the analytic structure of the spectral density.

Expressing the $A$'s and $B$'s in eqs.~(\ref{aiom})
and~(\ref{a-and-b})
in terms of $r$ and using \Eq{eq-mR} to eliminate $N$ gives
\begin{eqnarray}
   A^2 & = & F^2 \left( \sinh^2\mR  + (r+3)/2 \right),  \nonumber \\
   B_1^2 & = & F^2 \left( \sinh^2\mR  - (r+3)/2 \right), \nonumber \\
   B_2^2 & = & F^2 \left( \sinh^2\mR  + (r+3)/2 \right), \nonumber \\
   B_3^2 & = & F^2 \left( \sinh^2\mR  - (r+3)/6 \right), \nonumber \\
   B_4^2 & = & F^2 \left( \sinh^2\mR  + (r+3)/6 \right).
   \label{a-and-b-r}
\end{eqnarray}
\clearpage
\noindent
   We have now reduced our original four-dimensional integrals to one
 dimensional integrals from which we can find the analytic structure
 of $a_i(\omega)$ and therefore also of $\tilde{C}_i(\omega)$.
 In an infinite volume $a_i(\omega)$ is given by an integral
 of the form
\begin{eqnarray}
   a_i(\omega) & =& \frac{1}{F^2} \int_{-3}^3 dr \     \phi(r)
   \left( \frac{f_+(r)}{2  \sinh^2\mR + r +4 -\cos \omega } \right. \nonumber\\
  & &
  -\, \left.  \frac{f_-(r)}{2  \sinh^2\mR + r +4 +\cos \omega } \right) \,.
   \label{a-inf}
\end{eqnarray}
$f_+(r)$ and $f_-(r)$ are analytic (and real) in $(-3,3)$. The ensuing
singularity structure is sketched in \Fig{fig-cuts} and discussed in the
main text.

\end{document}